\PassOptionsToPackage{unicode}{hyperref}
\PassOptionsToPackage{hyphens}{url}
\documentclass[
]{article}
\usepackage{amsmath,amssymb}
\usepackage{lmodern}
\usepackage{iftex}
\ifPDFTeX
  \usepackage[T1]{fontenc}
  \usepackage[utf8]{inputenc}
  \usepackage{textcomp} 
\else 
  \usepackage{unicode-math}
  \defaultfontfeatures{Scale=MatchLowercase}
  \defaultfontfeatures[\rmfamily]{Ligatures=TeX,Scale=1}
\fi
\IfFileExists{upquote.sty}{\usepackage{upquote}}{}
\IfFileExists{microtype.sty}{
  \usepackage[]{microtype}
  \UseMicrotypeSet[protrusion]{basicmath} 
}{}
\makeatletter
\@ifundefined{KOMAClassName}{
  \IfFileExists{parskip.sty}{%
    \usepackage{parskip}
  }{
    \setlength{\parindent}{0pt}
    \setlength{\parskip}{6pt plus 2pt minus 1pt}}
}{
  \KOMAoptions{parskip=half}}
\makeatother
\usepackage{xcolor}
\usepackage{longtable,booktabs,array}
\usepackage{calc} 
\usepackage{etoolbox}
\makeatletter
\patchcmd\longtable{\par}{\if@noskipsec\mbox{}\fi\par}{}{}
\makeatother
\IfFileExists{footnotehyper.sty}{\usepackage{footnotehyper}}{\usepackage{footnote}}
\makesavenoteenv{longtable}
\usepackage{graphicx}
\makeatletter
\def\maxwidth{\ifdim\Gin@nat@width>\linewidth\linewidth\else\Gin@nat@width\fi}
\def\maxheight{\ifdim\Gin@nat@height>\textheight\textheight\else\Gin@nat@height\fi}
\makeatother
\setkeys{Gin}{width=\maxwidth,height=\maxheight,keepaspectratio}
\makeatletter
\def\fps@figure{htbp}
\makeatother
\ifLuaTeX
  \usepackage{selnolig}  
\fi

\IfFileExists{bookmark.sty}{\usepackage{bookmark}}{\usepackage{hyperref}}
\IfFileExists{xurl.sty}{\usepackage{xurl}}{} 
\hypersetup{
  hidelinks,
  pdfcreator={LaTeX via pandoc}}
\usepackage{booktabs}
\usepackage{array}
\usepackage{tabularx}
\usepackage{ragged2e}
\author{}
\date{}

\begin{document}

\begin{center}
\textbf{Physically Unclonable Functions for Secure IoT Authentication\\
and Hardware-Anchored AI Model Integrity}

\vspace{0.5cm}

Maryam Taghi Zadeh$^{1}$ and Mohsen Ahmadi$^{1}$

\vspace{0.3cm}

$^{1}$ Department of Electrical and Computer Science, Florida Atlantic University, Boca Raton, FL, USA

\vspace{0.2cm}

\textit{Corresponding authors:} \\
\href{mailto:mahmadi2021@fau.edu}{mahmadi2021@fau.edu}, 
\href{mailto:mtaghizadeh2023@fau.edu}{mtaghizadeh2023@fau.edu}
\end{center}
\textbf{Abstract}

The rapid integration of artificial intelligence (AI) into Internet of
Things (IoT) and edge computing systems has intensified the need for
robust, hardware-rooted trust mechanisms capable of ensuring device
authenticity and AI model integrity under strict resource and security
constraints. This survey reviews and synthesizes existing literature on
hardware-rooted trust mechanisms for AI-enabled IoT systems. It
systematically examines and compares representative trust anchor
mechanisms, including Trusted Platform Module (TPM)--based measurement
and attestation, silicon and FPGA-based Physical Unclonable Functions
(PUFs), hybrid container-aware hardware roots of trust, and
software-only security approaches. The analysis highlights how
hardware-rooted solutions generally provide stronger protection against
physical tampering and device cloning compared to software-only
approaches, particularly in adversarial and physically exposed
environments, while hybrid designs extend hardware trust into runtime
and containerized environments commonly used in modern edge deployments.
By evaluating trade-offs among security strength, scalability, cost, and
deployment complexity, the study shows that PUF-based and hybrid trust
anchors offer a promising balance for large-scale, AI-enabled IoT
systems, whereas software-only trust mechanisms remain insufficient in
adversarial and physically exposed settings. The presented comparison
aims to clarify current design challenges and guide future development
of trustworthy AI-enabled IoT platforms.


\textbf{1. Introduction}

The rapid proliferation of network-connected devices has driven
unprecedented growth of the Internet of Things (IoT) ecosystem. Smart
devices including mobile platforms, wireless sensors, wearables,
intelligent vehicles, and embedded control units are now widely deployed
across healthcare, transportation, industrial automation, smart cities,
and critical infrastructure. This large-scale deployment has led to
massive data generation and exchange, transforming how information is
sensed, processed, and transmitted {[}1-5{]}. As IoT systems mature,
they increasingly incorporate edge artificial intelligence (AI) to
enable local data processing, real-time decision-making, and autonomous
operation. Despite these advancements, ensuring secure and reliable data
handling in IoT remains a major challenge {[}6{]}. In addition, many IoT
applications must satisfy strict real-time constraints, requiring
low-latency processing and timely decision-making to ensure reliable
system performance {[}7{]}. IoT devices are typically distributed,
resource-constrained, and wirelessly connected, and they often operate
in untrusted or physically exposed environments. These conditions allow
adversaries to gain physical or logical access to hardware platforms,
making hardware-level vulnerabilities a critical security concern,
especially when sensitive or confidential data is involved {[}7,10{]}.

Exposure to adversarial environments has intensified hardware-focused
threats such as device theft, counterfeiting, cloning, and supply-chain
manipulation, which are widely recognized as core IoT security
challenges in systematic security surveys {[}15{]}. These attacks
include unauthorized cloning, overproduction, substitution, or tampering
of integrated circuits (ICs), intellectual property (IP) cores, and
complete embedded devices {[}11,12{]}. Rogue components introduced into
legitimate systems undermine device authenticity, reliability, and
trust, while also causing significant economic losses and supply-chain
disruptions {[}13,14{]}. As a result, defending against hardware theft
and counterfeiting has become an urgent requirement for IoT security. In
large-scale IoT deployments, adversaries can intercept, replace, or
physically tamper with individual devices or subsystems, leading to data
leakage, privilege escalation, fault injection, denial-of-service
attacks, or even network-wide failures. Existing authentication
mechanisms often rely on chip-level identification, which does not
guarantee system-level integrity {[}16-19{]}. A device may contain
authentic chips while still being compromised through partial
modification or malicious substitution. Traditional hardware
authentication approaches primarily rely on cryptographic techniques
that store secret keys in non-volatile memory {[}20,21{]}. While
effective in theory, these methods are vulnerable to key extraction via
reverse engineering or physical attacks and impose substantial
computational and energy overhead {[}22,23{]}. Such limitations make
software-centric cryptographic solutions less suitable for
resource-constrained IoT devices and highlight the need for lightweight,
hardware-rooted security primitives {[}24-25{]}.

The growing deployment of AI models on IoT devices further expands the
attack surface. AI model parameters trained in cloud environments are
often transmitted to edge devices, where insufficient protection can
expose them to tampering, replacement, or unauthorized modification.
Attacks such as model tampering, malicious model replacement, and
backdoor insertion can compromise inference behavior in subtle yet
dangerous ways. These risks are amplified in IoT environments, where
devices are typically physically accessible and operate under limited
security controls. Consequently, ensuring the integrity,
confidentiality, and authenticity of AI model parameters has become a
critical requirement for secure edge intelligence {[}26,27{]}.

\includegraphics[width=4.81836in,height=2.6075in]{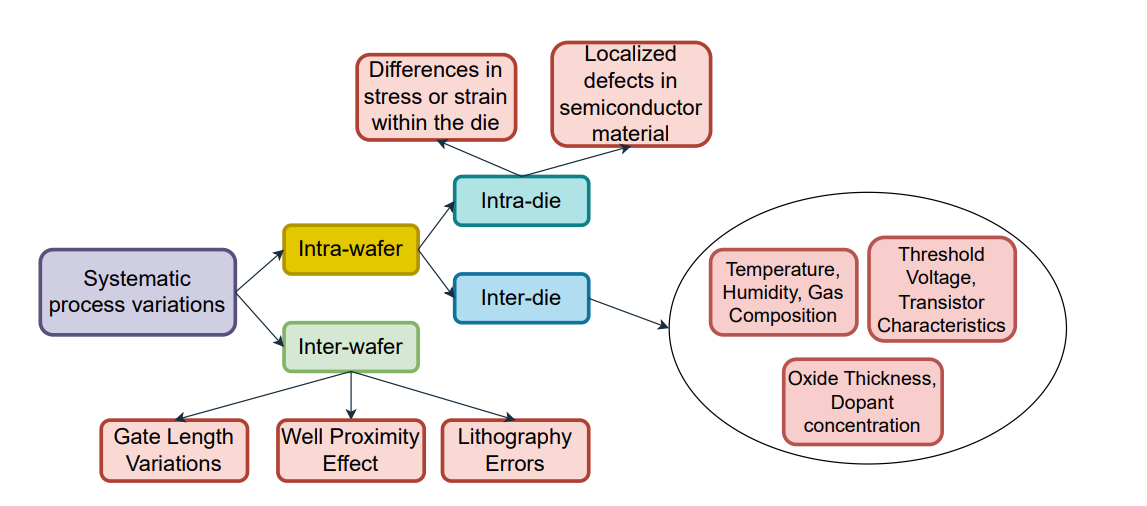}

\textbf{Figure 1}: Types of systematic process variations in
semiconductor devices

Physical Unclonable Functions (PUFs) have emerged as a promising
hardware-based solution for IoT authentication and AI model protection.
By exploiting inherent manufacturing variations in semiconductor
devices, PUFs generate unique, device-specific fingerprints without
storing secret keys in memory. This intrinsic hardware identity provides
a strong foundation for secure device authentication, hardware-rooted
trust, and binding AI models to legitimate IoT platforms {[}28,29{]}.
Figure 1 illustrates the main categories of systematic process
variations---inter-wafer, intra-wafer, inter-die, and intra-die---and
their underlying causes, including lithography errors, gate-length
variations, material defects, and environmental and device-level
parameter fluctuations.

Existing approaches for protecting AI parameters in IoT systems
generally fall into two categories: cryptography-based encryption
schemes and hardware-based authentication mechanisms {[}7,8,20,30{]}.
Cryptographic methods rely on symmetric or public-key encryption to
secure communication channels and verify device identities, while
techniques such as hashing and message authentication codes ensure data
integrity. Although effective in principle, these approaches remain
vulnerable to key leakage during storage and distribution, and their
computational and hardware overhead can be prohibitive for
resource-constrained IoT devices. Hardware-based security primitives,
particularly PUFs, offer an alternative by exploiting intrinsic physical
properties of devices that cannot be precisely replicated, even by the
original manufacturer {[}31,32{]}.

Typically, a PUF operates through a challenge--response mechanism in
which responses depend on the device's physical characteristics
{[}32,33{]}. Even identical PUF designs fabricated using the same
process produce distinct responses due to inherent process variations,
making PUFs particularly attractive for lightweight and low-power IoT
applications. PUFs have been widely applied in RFID systems, secure
communication protocols, IP protection, cryptographic key generation,
and device authentication. With the rapid growth of edge-AI,
hardware-rooted trust has become increasingly important in large-scale
and distributed IoT deployments. However, deploying machine learning
inference directly on edge devices introduces new security risks,
including model extraction, unauthorized modification, malicious model
replacement, and device impersonation. PUF-based trust anchors help
mitigate these threats by enabling unclonable device authentication,
binding AI model parameters to legitimate hardware, supporting secure
boot and runtime integrity checks, and establishing trusted
communication channels between edge nodes and remote servers.

Among various PUF implementations, Static Random Access Memory (SRAM)
PUFs have received significant attention due to their cost efficiency,
scalability, and compatibility with existing digital circuits
{[}34,35{]}. SRAM PUFs exploit the power-up behavior of SRAM cells,
which is influenced by slight mismatches in transistor threshold
voltages introduced during fabrication. Upon power-up, each SRAM cell
settles into a preferred state, producing a unique and repeatable
response, where the memory address serves as the challenge and the
power-up value as the response {[}36,27{]}. SRAM PUFs can be integrated
into standard SRAM arrays already present in most digital systems,
minimizing area and power overhead while enabling the generation of
large numbers of response bits. However, SRAM PUFs are sensitive to
noise and environmental variations, leading to response instability in
the form of bit flips. Factors such as temperature fluctuations, voltage
variations, and device aging can degrade response reliability and reduce
entropy over time.

This review presents a structured and reliability-focused synthesis of
SRAM-based Physical Unclonable Functions (PUFs), integrating prior
studies across fabrication processes, reliability behavior, and
data-driven authentication. First, the paper examines
fabrication-induced process variations in SRAM chips and their impact on
PUF behavior, highlighting how fine-grained manufacturing disparities
such as random dopant fluctuation, line-edge roughness, and
intra-/inter-die variations simultaneously enable device uniqueness and
introduce reliability challenges often overlooked in existing surveys.
Second, the review consolidates research on SRAM PUF reliability
degradation, emphasizing bit-flip behavior, neighboring-cell effects,
and sensitivity to voltage, temperature, and aging. By framing
instability as an evolving pattern rather than isolated noise, the paper
clarifies the implications of long-term degradation on response
consistency and entropy. Third, this work organizes and evaluates
machine-learning-based approaches for SRAM PUF analysis, including
manufacturer identification and new-versus-aged chip classification. The
review highlights commonly used models and their suitability for
numerical PUF features, demonstrating how learning-based methods
complement traditional hardware-centric security analysis. The paper
discusses the practical relevance of SRAM PUF classification for device
authentication, trust assessment, and quality control, underscoring its
importance for secure IoT and embedded systems.

Beyond the SRAM-focused analysis, this survey further extends the
discussion to a broader comparison of hardware-rooted trust anchor
mechanisms for AI-enabled IoT systems. To enable a structured
comparison, this survey evaluates trust anchor mechanisms based on the
following criteria: (i) underlying trust anchor type, (ii) secret/key
storage model, (iii) scalability in large-scale IoT deployments, (iv)
suitability for resource-constrained devices, (v) support for AI model
integrity, and (vi) implementation overhead and limitations. The
comparison synthesizes findings from representative recent literature
and organizes them into a unified analytical framework.

\textbf{2. Related Work}

Arul Selvan et al. {[}30{]} presented a secure bootloader framework for
embedded systems that establishes a hardware-based chain of trust using
cryptographic validation, ensuring that only authenticated firmware is
executed and effectively preventing firmware tampering, rollback
attacks, and unauthorized code modification throughout the device
lifecycle. Siyal et al. {[}31{]} introduced a blockchain-based
supply-chain provenance framework that binds embedded PUF-generated
hardware identities to NFTs, enabling tamper-evident product
authentication and decentralized verification, with edge and 6G networks
supporting low-latency validation.

Kasimatris et al. {[}32{]} proposed a decentralized IoT device identity
framework that integrates PUFs with blockchain-based Soulbound Tokens to
achieve hardware-anchored authentication and secure lifecycle
traceability across supply-chain stages, demonstrating practical
feasibility with moderate computational overhead. Tran et al. {[}33{]}
proposed a lightweight end-to-end security protocol for IoT devices that
combine CRP-based PUFs and a TRNG to eliminate nonvolatile key storage,
achieving ECC-compatible authentication with reduced computational
overhead and strong resistance to physical and implementation-level
attacks. Sarkar et al. {[}35{]} presented a comprehensive survey of
secure communication in drone networks, systematically classifying
lightweight encryption and key management techniques across UAV protocol
layers, with emphasis on energy efficiency, post-quantum readiness, and
scalable security architectures for dynamic aerial environments.

Lai et al. {[}36{]} proposed an authentication and key agreement scheme
for in-vehicle networks that leverages SRAM-based PUFs to secure CAN bus
communication and ECU authentication, effectively resisting replay and
physical attacks while maintaining low latency and moderate overhead.
Venugopal et al. {[}37{]} proposed a blockchain-based provenance and
data-integrity framework for environmental electrochemical sensor
networks, combining edge processing with cryptographic commitments to
enable end-to-end verifiable analytics, efficient calibration
validation, and strong tamper detection without centralized trust. Khan
et al. {[}38{]} investigated probing attacks on advanced IC packaging
and proposed a standardized vulnerability assessment framework that
quantitatively evaluates security risks in heterogeneous and 2.5D/3D
packaging technologies, highlighting the need for new protection metrics
alongside performance gains. Zheng et al. {[}39{]} proposed a
lightweight PUF-based secure group communication protocol for
low-altitude networks that supports dynamic group membership, eliminates
long-term key storage, and achieves efficient group key renewal with low
computational and communication overhead. Casado-GalÃ¡n et al. {[}40{]}
analyzed electromagnetic side-channel leakage in Ring Oscillator PUF
implementations, identified key vulnerability sources, and proposed
hardware-level countermeasures that mitigate information leakage with
moderate area overhead. Pawlik et al. {[}41{]} presented a systematic
review of cybersecurity challenges in electric vehicle charging
infrastructure, categorizing cyber-attacks, intrusion and anomaly
detection methods, and authentication mechanisms to assess risks and
mitigation strategies for V2G-enabled systems. Samanta et al. {[}42{]}
analyzed the impact of temperature variations on SRAM-based PUFs for
low-cost embedded systems, showing that most cells retain stable
power-up behavior across a wide temperature range while preserving
inter-device uniqueness.

Table 1: Summary of Physical Unclonable Function (PUF)--Based Methods
for IoT and IIoT Security

\begin{longtable}[]{@{}
  >{\raggedright\arraybackslash}p{(\columnwidth - 10\tabcolsep) * \real{0.1453}}
  >{\raggedright\arraybackslash}p{(\columnwidth - 10\tabcolsep) * \real{0.0643}}
  >{\raggedright\arraybackslash}p{(\columnwidth - 10\tabcolsep) * \real{0.0702}}
  >{\raggedright\arraybackslash}p{(\columnwidth - 10\tabcolsep) * \real{0.1721}}
  >{\raggedright\arraybackslash}p{(\columnwidth - 10\tabcolsep) * \real{0.2310}}
  >{\raggedright\arraybackslash}p{(\columnwidth - 10\tabcolsep) * \real{0.3171}}@{}}
\toprule()
\begin{minipage}[b]{\linewidth}\raggedright
\textbf{Author(s)}
\end{minipage} & \begin{minipage}[b]{\linewidth}\raggedright
\textbf{Ref.}
\end{minipage} & \begin{minipage}[b]{\linewidth}\raggedright
\textbf{Year}
\end{minipage} & \begin{minipage}[b]{\linewidth}\raggedright
\textbf{Method}
\end{minipage} & \begin{minipage}[b]{\linewidth}\raggedright
\textbf{Aim}
\end{minipage} & \begin{minipage}[b]{\linewidth}\raggedright
\textbf{Result}
\end{minipage} \\
\midrule()
\endhead
Cheon et al. & {[}54{]} & 2026 & Graphene adlayer morphology--based PUF
on microfaceted Cu substrates & To exploit spatial randomness and
thickness nonuniformity of graphene as a physical entropy source &
Achieved strong PUF metrics; deep learning--based classification
confirmed high distinguishability and suitability for secure hardware
authentication \\
Zhang et al. & {[}52{]} & 2026 & Self-assembled optical PUF using
microsphere-filled silicon microholes & To develop a
fabrication-compatible and unclonable PUF for IIoT authentication and
secure communication & Demonstrated strong unclonability through random
self-assembly; enabled optical CRP extraction and supported
anti-counterfeiting and encrypted IIoT communication \\
Wang et al. & {[}51{]} & 2026 & Review of architected nanomaterial-based
optical PUFs (OPUFs) & To survey how engineered nanomaterials enhance
optical PUF performance & Showed significant improvements in entropy,
encoding capacity, robustness, and security; identified scalability and
standardization as open challenges \\
Chang et al. & {[}49{]} & 2026 & On-chip nonlinear optical PUF using
thin-film lithium niobate arrays & To design an integrated optical PUF
exploiting nonlinear scattering for IoT authentication & Demonstrated
high unpredictability and unclonability with a CRP space of
approximately 10â¶ and strong resistance to modeling and link attacks \\
Wei et al. & {[}48{]} & 2026 & Wearable-compatible all-optical PUF with
hybrid deep learning & To enable secure authentication for wearable
devices using deep-learning-enhanced optical PUFs & Achieved 91.7\%
authentication accuracy and up to 98.75\% forgery detection, suitable
for wearable and IoT security \\
Boghban-Bousari et al. & {[}53{]} & 2025 & Pre-stressed OTFT-based
electronic PUF & To evaluate pre-stressed OTFT current variability as a
reliable entropy source & Demonstrated high reproducibility
(\textasciitilde0.99) with balanced uniformity and uniqueness
(\textasciitilde0.52 and \textasciitilde0.50), improving OTFT-based PUF
reliability for cryptographic applications \\
Cao et al. & {[}50{]} & 2025 & Colloidal nanowire-based AI-resilient PUF
with triple-key authentication & To resist AI-driven modeling attacks
using nanoscale structural randomness & Maintained prediction accuracy
below 62\% under evaluated AI attacks \\
Alhamarneh, Singh & {[}46{]} & 2024 & Systematic survey of PUF-based IoT
authentication protocols & To analyze PUF-based security mechanisms and
cloning challenges in IoT & Proposed the PUF3S-ML framework and
highlighted unresolved issues in cloning resistance, deployment, and
protocol integration \\
Al-Meer, Al-Kuwari & {[}44{]} & 2023 & Comprehensive survey of PUF
architectures and IoT protocols & To evaluate PUFs as lightweight
alternatives to conventional cryptography & Concluded that PUFs offer
low-cost, low-power security while facing challenges related to
reliability, attacks, and standardization \\
Shan et al. & {[}47{]} & 2021 & PUF-based verifiable data stream
transmission for IIoT & To secure industrial sensor data streams under
resource constraints & Demonstrated improved data integrity protection
and communication efficiency compared to traditional cryptographic
schemes \\
Zhang et al. & {[}45{]} & 2019 & PUF-based anonymous authentication with
Merkle hash trees & To provide privacy-preserving, multi-access
authentication for IIoT & Reduced authentication overhead while ensuring
anonymity and efficiency for resource-constrained IIoT devices \\
\bottomrule()
\end{longtable}

Table 1 presents a chronological comparison of representative Physical
Unclonable Function (PUF)--based studies addressing security challenges
in IoT and Industrial IoT (IIoT) environments. The reviewed works span
electronic, optical, nanomaterial-based, and AI-resilient PUF
architectures, highlighting the evolution from conventional electronic
PUFs toward advanced optical and material-driven solutions. Earlier
studies primarily focused on lightweight authentication and data
integrity for resource-constrained IIoT devices, while more recent works
emphasize resistance to modeling attacks, deep-learning-based forgery
detection, and scalability for emerging applications such as wearable
devices and hardware anti-counterfeiting. Survey papers systematically
summarize architectural trends and identify persistent challenges,
including reliability, environmental robustness, standardization, and
large-scale deployment.

\textbf{ 3- Internet of Things (IoT) and Cyber-Physical Systems (CPS) Security Challenges}

The evolution of Industry 4.0 has driven the widespread adoption of
Cyber-Physical Systems (CPS) as a fundamental component of modern
Internet of Things (IoT) environments. CPS-enabled smart factories,
industrial automation platforms, intelligent transportation systems, and
connected healthcare infrastructures increasingly depend on
computer-controlled mechanical processes that were traditionally
performed by humans {[}55,56{]}. These systems integrate physical
components with computational intelligence and networking technologies,
enabling machines, devices, and services to communicate through the
Internet and cloud platforms. While this paradigm significantly improves
efficiency, flexibility, and productivity, it also introduces
considerable security risks {[}57,58{]}. A Cyber-Physical System can be
defined as an embedded network that monitors and controls physical
processes using computer-based algorithms. CPS combines sensors,
aggregators, and actuators to acquire real-world data, process it
digitally, and execute control actions in real time {[}59,60{]}. These
components are typically connected via wired or wireless networks,
allowing remote monitoring, decentralized decision-making, and adaptive
system behavior. CPS shares many characteristics with IoT, as both
involve interconnected devices interacting with physical environments.
In practice, CPS serves as the operational backbone of IoT applications
by enabling digital control over physical processes {[}61{]}.
CPS-enabled IoT systems are deployed across diverse domains, including
industrial control systems, smart grids, oil refineries, water treatment
plants, medical devices, robotics, automotive systems, supply chains,
and smart cities {[}62{]}. The large scale and heterogeneity of these
systems complicate management and security, especially when devices
originate from different manufacturers and rely on diverse hardware and
software stacks. As connectivity increases, CPS components that were
once isolated become accessible through open networks, significantly
expanding the attack surface {[}63-65{]}.

Security challenges in IoT and CPS environments arise primarily from
their distributed architecture, continuous operation, and deployment in
partially trusted or untrusted settings. Sensitive data collected by
sensors such as industrial process parameters, medical information, or
infrastructure status must be protected from unauthorized access and
manipulation. Interception or disruption of communication signals can
compromise privacy and system reliability, reducing trust in IoT
technologies and limiting their adoption {[}66,67{]}.

Several underlying factors contribute to CPS security vulnerabilities.
Traditional CPS architecture relied on physical and network isolation
for protection, but modern systems emphasize connectivity, exposing them
to external cyber threats {[}68{]}. The widespread use of wireless
communication, cloud services, and open protocols has shifted attacks
from internal sources to Internet-based adversaries. Moreover, CPS
systems integrate heterogeneous components, often supplied by
third-party vendors, each introducing potential weaknesses.
Vulnerabilities in one component can propagate across the system,
amplifying overall risk {[}69,70{]}. Vulnerabilities manifest at
multiple layers of CPS-enabled IoT systems. Communication
vulnerabilities stem from reliance on protocols such as TCP/IP, which
were not designed for real-time or safety-critical control. Operating
system vulnerabilities are common, particularly in real-time operating
systems that lack fine-grained access control. Software vulnerabilities
further exacerbate security risks when control applications or
programmable logic controllers lack integrity verification mechanisms,
enabling malicious code injections {[}71{]}.

\includegraphics[width=5.5in,height=4.36389in]{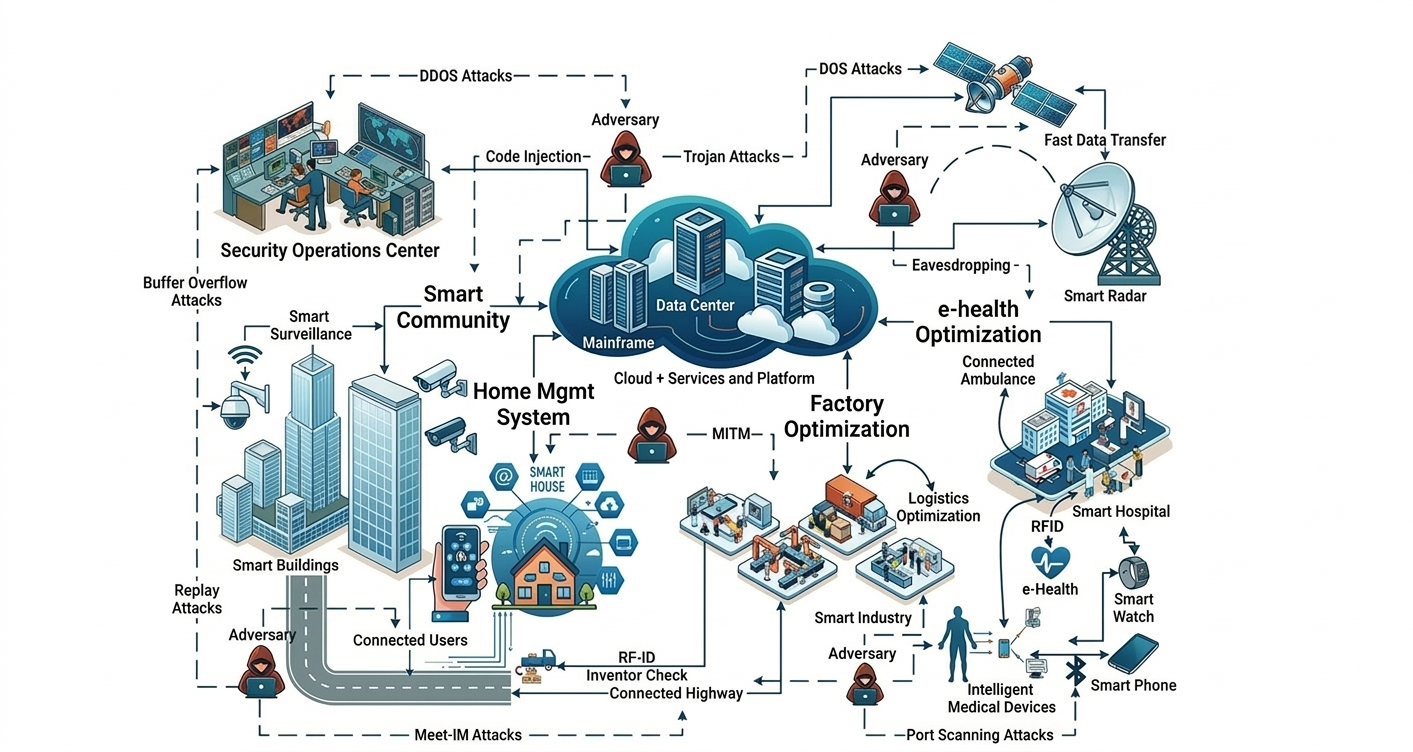}

\textbf{Figure 2}: IoT and cyber-physical systems security challenges in
smart community and cloud-based infrastructures

Figure 2 illustrates a high-level architecture of a smart community
ecosystem that integrates smart homes, smart buildings, industrial
automation, e-health systems, and cloud-based services. It also
summarizes representative cyber-physical threat categories such as DDoS,
man-in-the-middle (MITM), replay, and eavesdropping that may target
different layers of the system, from edge devices and communication
links to cloud platforms. The attack labels are intended as illustrative
examples of common threat types and are not exhaustive {[}72,73{]}.
These weaknesses enable both physical and cyber-attacks, including
device impersonation, tampering, social engineering, replay attacks,
data injection, malware deployment, and denial-of-service attacks.
Unlike conventional IT systems, successful attacks on CPS can directly
impact physical processes, resulting in equipment damage, safety
hazards, service outages, and significant financial losses {[}74,75{]}.

\textbf{Table 2:} Security Challenges in IoT and CPS Environments

\begin{longtable}[]{@{}
  >{\raggedright\arraybackslash}p{(\columnwidth - 2\tabcolsep) * \real{0.2638}}
  >{\raggedright\arraybackslash}p{(\columnwidth - 2\tabcolsep) * \real{0.7362}}@{}}
\toprule()
\begin{minipage}[b]{\linewidth}\raggedright
\textbf{Aspect}
\end{minipage} & \begin{minipage}[b]{\linewidth}\raggedright
\textbf{Description}
\end{minipage} \\
\midrule()
\endhead
CPS Connectivity & Internet-enabled CPS increases exposure to external
cyber threats \\
System Heterogeneity & Integration of third-party hardware and software
introduces vulnerabilities \\
Data Privacy & Sensitive sensor and user data may be intercepted or
leaked \\
Communication Security & Reliance on standard protocols exposes systems
to interception \\
Software Vulnerabilities & Lack of authentication and code integrity
mechanisms \\
Physical Attacks & Device impersonation, tampering, and side-channel
leakage \\
Cyber Attacks & Eavesdropping, replay, malware, and denial-of-service
attacks \\
System Impact & Service disruption, physical damage, financial loss, and
safety risks \\
\bottomrule()
\end{longtable}

The diversity of IoT and CPS applications makes it difficult to deploy
uniform security solutions. Different domains impose distinct
requirements for latency, reliability, and safety, requiring security
mechanisms that balance protection with efficiency. While cryptographic
approaches remain essential, their computational overhead, key
management complexity, and vulnerability to physical compromise limit
their effectiveness in resource-constrained environments. Consequently,
developing lightweight, scalable trust mechanisms that address both
cyber and physical threats remains a critical research challenge for
CPS-enabled IoT systems (see Table 2).

\textbf{4- Hardware Security and Physical Roots of Trust}

Modern embedded and IoT systems increasingly rely on hardware-based
security mechanisms to establish trustworthy operation in environments
that are both constrained and physically exposed. Hardware-security
systems aim to guarantee reliable device identification, authentication,
and secure communication by embedding security directly into circuit and
architectural design {[}76,77{]}. Central to this approach is the
hardware root of trust, which provides an initial trusted anchor from
which security can be propagated through firmware, operating systems,
and communication protocols {[}78{]}. Traditional roots of trust are
often implemented using secret keys stored in non-volatile memory or
programmed via electronic fuses. While conceptually straightforward,
these approaches suffer from inherent vulnerabilities, as stored secrets
can be extracted through invasive or non-invasive physical attacks such
as probing, optical inspection, or reverse engineering {[}79,80{]}. In
addition, secure storage mechanisms introduce non-negligible areas,
power, and cost overhead, making them less attractive for low-power IoT
and edge devices. These limitations have driven significant interest in
security primitives that derive trust from physical properties rather
than permanent key storage {[}81-85{]}.

Hardware security primitives encompass a collection of circuit-level
building blocks that support secure operation {[}86{]}. These include
mechanisms for key generation, random number generation, encryption and
decryption, and authentication. In practice, the choice of primitives
and protocols is strongly influenced by energy constraints {[}82,83{]}.
Public-key cryptography, although essential for initial trust
establishment, incurs substantially higher energy costs compared to
sensing, data processing, or symmetric-key encryption. Consequently,
hardware-secure architecture typically minimizes the use of public-key
operations and relies primarily on lightweight symmetric cryptography
once trust has been established. This explicit trade-off between energy
efficiency and security strength is a defining feature of hardware
security design {[}84-86{]}. Supporting primitives such as random number
generators and error management mechanisms are also essential. Random
numbers are required for nonces, session keys, and initialization
vectors, while error correction becomes necessary when security
primitives rely on noisy physical effects. Together, these components
form the foundation of secure hardware systems by anchoring trust in
silicon-level behavior rather than software abstractions.

\textbf{5-Physical Unclonable Functions (PUFs)}

Physical Unclonable Functions (PUFs) represent a prominent class of
hardware security primitives that exploit uncontrollable manufacturing
variations inherent in semiconductor fabrication {[}87{]}. Instead of
storing secrets in memory, PUFs generate device-specific responses on
demand by measuring physical properties of the circuit. Because these
responses are not permanently stored and only exist during operation,
PUFs are naturally more resistant to physical extraction attack. The
fundamental principle behind PUF operation is the amplification of
random within-die variations such as threshold voltage mismatch, delay
differences, or current imbalance while suppressing environmental noise
and systematic variation. The resulting responses serve as a unique
silicon fingerprint, enabling device identification and authentication
even among chips fabricated using the same process and layout {[}88{]}.
In challenge--response-based protocols, a verifier authenticates a
device by comparing its response to a previously enrolled reference,
making cloning or prediction extremely difficult without access to the
original hardware.

\includegraphics[width=5.26667in,height=3.10833in]{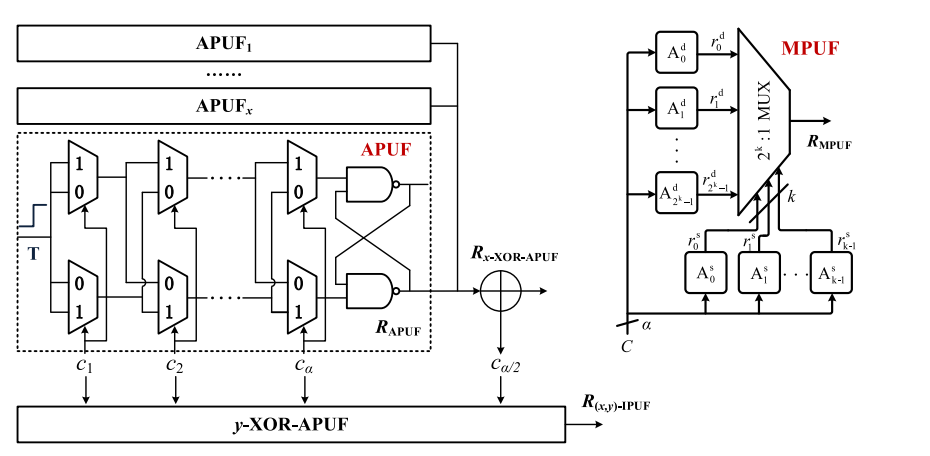}

\textbf{Figure 3}: Architecture of representative strong PUF
constructions, including the XOR Arbiter PUF (y-XOR-APUF) and the
multiplexer-based MPUF, illustrating how multiple PUF instances and
challenge-controlled selection are combined to increase response
complexity and enhance resistance against direct prediction.

Figure 3 illustrates two representative strong PUF architectures: the
y-XOR Arbiter PUF (left) and the multiplexer-based PUF (MPUF) (right).
In the y-XOR-APUF, multiple arbiter PUF instances are evaluated under
the same challenge, and their outputs are XOR-combined to increase
response complexity and reduce predictability. In the MPUF design, the
challenge controls a multiplexer that selects among multiple internal
response paths, creating a more complex challenge--response relationship
and improving resistance to direct modeling. PUFs are commonly
classified according to their functionality and implementation. Weak
PUFs support a limited number of challenges--response pairs and are
typically used for key generation or unique identifiers {[}89{]}. These
applications require high response stability, as even small error rates
can lead to cryptographic failures. As a result, weak PUFs often rely on
post-processing techniques such as error correction and unstable-bit
suppression. Strong PUFs, in contrast, support a large number of
challenges--response pairs and are primarily used for authentication.
While strong PUFs can tolerate moderate noise, they are generally more
vulnerable to modeling attacks, where adversaries attempt to learn the
PUF behavior from observed challenge--response pairs {[}89{]}.

\includegraphics[width=4.5in,height=2.01944in]{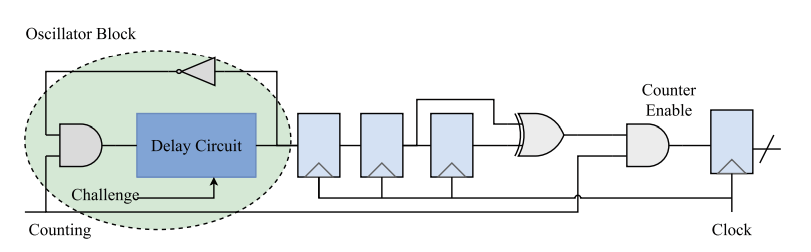}

\textbf{Figure 4}: Self-oscillating loop architecture of a
Ring-Oscillator Physical Unclonable Function (RO-PUF)

Figure 4 illustrates the operating principle of a Ring-Oscillator
Physical Unclonable Function (RO-PUF) based on a self-oscillating loop.
The core of the design is an oscillator block composed of a delay
circuit whose configuration is controlled by an external challenge
input. The delay circuit introduces path-dependent propagation delays
that are influenced by intrinsic manufacturing variations.

\includegraphics[width=4.5in,height=2.90069in]{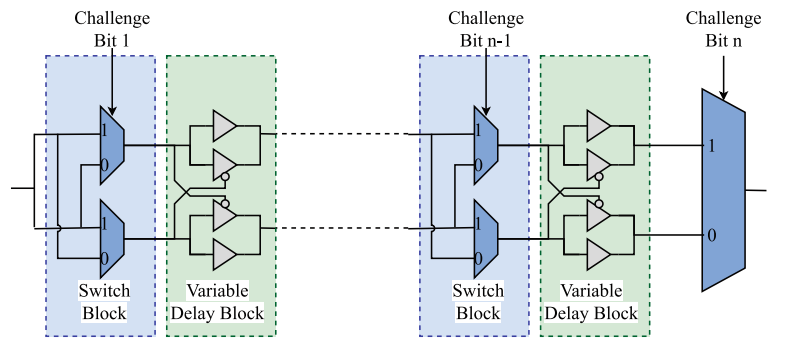}

\textbf{Figure 5}: Non-monotonic delay circuit used in delay-based PUF
architectures.

Figure 5 illustrates a non-monotonic delay circuit commonly used in
delay-based PUFs, particularly in advanced Arbiter PUF and Ring
Oscillator PUF designs. The architecture consists of a sequence of
switch blocks followed by variable delay blocks, where each stage is
controlled by a corresponding challenge bit. Another important distinction is between intrinsic and extrinsic PUFs. Intrinsic PUFs leverage variations already present in standard circuit
components, such as SRAM cells or logic gates, and therefore require
minimal additional hardware {[}90{]}. Extrinsic PUFs introduce dedicated
structures designed specifically to enhance entropy. Although extrinsic
designs may provide stronger randomness, they often incur higher area
and power overhead, which can limit their applicability in constrained
systems.

\textbf{6-PUF Performance Metrics and Evaluation}

The effectiveness of a PUF is typically assessed using three core
metrics: reliability, uniqueness, and randomness. Reliability describes
the ability of a PUF to reproduce the same response when the same
challenge is applied repeatedly under varying environmental conditions.
Since temperature fluctuations, voltage variation, aging, and
measurement noise can all affect physical behavior, maintaining high
reliability is essential for practical deployment. Uniqueness measures
how well PUF distinguishes between different devices fabricated using
the same technology. Ideally, responses from different chips should
differ significantly when evaluated under the same challenge, minimizing
the probability of collisions {[}91{]}. This property is commonly
quantified using inter-device Hamming distance, with values near the
midpoint of the response length indicating strong uniqueness.

Randomness evaluates whether PUF responses are unbiased and
unpredictable. A response that consistently favors one value indicates
structural bias and weakens security by enabling prediction or modeling
attacks {[}92{]}. High entropy and balanced output distributions are
therefore essential for unclonability. These metrics are inherently
interdependent and improving one often impacts the others. Achieving
high uniqueness without sacrificing reliability, while maintaining
sufficient entropy, remains a central challenge in PUF design.

\textbf{7- Delay-Based and Racetrack PUF Architectures}

Delay-based PUFs form a major class of intrinsic PUFs that derive
responses from propagation delay variations caused by fabrication
randomness. This family includes Arbiter PUFs, Ring-Oscillator PUFs, and
Clock PUFs. Although their circuit implementations differ, they all rely
on comparing relative timing behavior to generate device-specific
responses. Arbiter PUFs compare the arrival times of two signals
propagating through challenge-configurable delay paths. While process
variation ensures uniqueness, conventional Arbiter PUFs can be
vulnerable to mathematical modeling if sufficient challenge--response
pairs are observed. Non-linear variants have been proposed to mitigate
this risk, though reliability remains sensitive to environmental
variation.

\begin{figure}[htbp]
\centering
\includegraphics[width=0.7\textwidth]{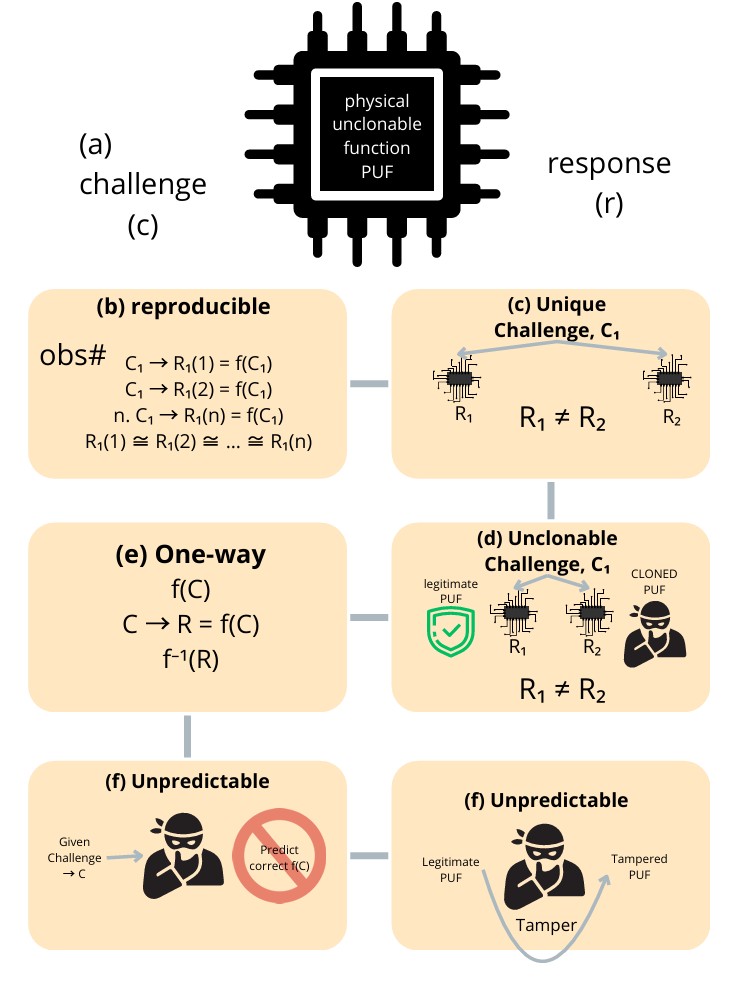}
\caption{Your caption here}
\label{fig:image6}
\end{figure}

\textbf{Figure 6}: Essential properties of Physical Unclonable
Functions, including reproducibility, uniqueness, unclonability, one-way
behavior, unpredictability, and tamper evidence.

Ring-Oscillator PUFs (RO-PUFs) embed delay elements within
self-oscillating loops, where device-specific frequency differences
arise from fabrication-induced delay variations {[}93{]}. In practical
deployments, RO-PUF implementations must address key challenges such as
measurement noise, environmental sensitivity (e.g., temperature and
voltage fluctuations), and correlated process variation. To enhance
robustness, prior studies commonly adopt compensation, calibration, and
oscillator-selection strategies (Figure 6). Clock PUFs, in
contrast, exploit clock skew in on-chip clock distribution networks,
under the assumption that environmental variations affect clock paths in
a relatively uniform manner {[}94{]}. By comparing clock transitions
across different paths, Clock PUFs can achieve high robustness with low
overhead and improved tamper resistance.

FPGA-based delay PUF implementations introduce additional challenges due
to limited control over placement and routing, which can amplify
systematic variation and reduce reproducibility. To mitigate these
effects, techniques such as balanced interconnect designs, XOR-based
response combination, and configurable ring oscillator structures have
been proposed to improve uniqueness and reliability while maintaining
area efficiency {[}95{]}. Although RO-PUFs remain the dominant
delay-based architecture, LC oscillator topologies with active inductors
provide an alternative design space, where both capacitive and inductive
variations may contribute to distinct and potentially richer PUF
response characteristics. Compact LC-based VCO designs using active
inductor structures have demonstrated wide tuning ranges {[}122{]},
suggesting promising opportunities for hybrid PUF architectures that
exploit different physical variation mechanisms beyond conventional ring
oscillator implementations.

\textbf{8- Classification of Physical Unclonable Functions}

Physical Unclonable Functions (PUFs) can be broadly classified according
to their fabrication characteristics and their security properties
{[}96{]} {[}43{]}. This classification helps organize the diverse range
of PUF architectures and clarifies their suitability for different
security applications. From a fabrication perspective, PUFs are
generally divided into silicon-based and non-silicon-based PUFs.
Non-silicon PUFs are constructed using materials outside conventional
semiconductor technologies. Examples include optical PUFs, paper PUFs,
magnetic PUFs, acoustic PUFs, and compact-disc (CD) PUFs {[}43{]}. These
approaches exploit physical randomness in macroscopic or
material-dependent phenomena and are often used in niche applications
where silicon integration is not required. While such PUFs can exhibit
strong uniqueness, they are typically less suitable for embedded or
large-scale IoT systems due to integration and scalability limitations.

Silicon-based PUFs, by contrast, exploit intrinsic and uncontrollable
variations that arise during semiconductor fabrication. These variations
include threshold voltage mismatches, delay differences, and memory cell
instabilities. Silicon PUFs are particularly attractive for integrated
systems because they can be implemented using standard CMOS processes
without requiring additional manufacturing steps. Based on the dominant
source of variation, silicon PUFs are commonly categorized into
delay-based and memory-based PUFs. Delay-based PUFs derive responses
from race conditions or frequency differences in signal propagation
paths, as observed in architectures such as Arbiter PUFs,
Ring-Oscillator PUFs, Clock PUFs, Glitch PUFs, and Interpose PUFs
{[}97{]}. Memory-based PUFs, on the other hand, exploit the startup
behavior or instability of memory elements, including SRAM PUFs,
Butterfly PUFs, latch-based PUFs, flip-flop PUFs, and DRAM PUFs.
Together, these silicon-based PUF families provide a versatile
foundation for lightweight hardware authentication and key generation in
resource-constrained IoT devices.

\textbf{Table 3.} Classification of Physical Unclonable Functions and
Representative Examples

\begin{longtable}[]{@{}
  >{\raggedright\arraybackslash}p{(\columnwidth - 6\tabcolsep) * \real{0.1985}}
  >{\raggedright\arraybackslash}p{(\columnwidth - 6\tabcolsep) * \real{0.1951}}
  >{\raggedright\arraybackslash}p{(\columnwidth - 6\tabcolsep) * \real{0.3168}}
  >{\raggedright\arraybackslash}p{(\columnwidth - 6\tabcolsep) * \real{0.2896}}@{}}
\toprule()
\begin{minipage}[b]{\linewidth}\raggedright
\textbf{Classification Criterion}
\end{minipage} & \begin{minipage}[b]{\linewidth}\raggedright
\textbf{Category}
\end{minipage} & \begin{minipage}[b]{\linewidth}\raggedright
\textbf{Description}
\end{minipage} & \begin{minipage}[b]{\linewidth}\raggedright
\textbf{Representative Examples}
\end{minipage} \\
\midrule()
\endhead
\textbf{Fabrication-Based} & Silicon PUFs & Exploit intrinsic
manufacturing variations in CMOS circuits; highly suitable for
integrated and embedded systems & SRAM PUF, Arbiter PUF, Ring-Oscillator
PUF, Clock PUF, Butterfly PUF, DRAM PUF \\
& Non-Silicon PUFs & Utilize physical randomness in non-silicon
materials or macroscopic structures; limited integration capability &
Optical PUF, Paper PUF, Magnetic PUF, Acoustic PUF, CD-PUF \\
\textbf{Variation Source} & Delay-Based PUFs & Derive responses from
race conditions or propagation delay differences in signal paths &
Arbiter PUF, RO-PUF, Clock PUF, Glitch PUF, Interpose PUF \\
& Memory-Based PUFs & Exploit startup behavior or instability of
volatile memory elements & SRAM PUF, Butterfly PUF, RS Latch PUF,
Flip-Flop PUF, DRAM PUF \\
\textbf{Security (CRP Space)} & Weak PUFs & Support a limited number of
challenge--response pairs; responses are kept on-chip & SRAM PUF,
RO-PUF, RS Latch PUF \\
& Strong PUFs & Support a large CRP space; used for challenge--response
authentication & Arbiter PUF, Bistable Ring PUF \\
\textbf{Implementation Style} & Intrinsic PUFs & Use existing circuit
components without dedicated PUF structures & SRAM PUF, Flip-Flop PUF \\
& Extrinsic PUFs & Introduce specialized circuitry to amplify randomness
& Arbiter PUF, Interpose PUF \\
\bottomrule()
\end{longtable}

Table 3 summarizes the classification of Physical Unclonable Functions
based on fabrication technology, underlying source of physical
variation, challenge--response space, and implementation style, along
with representative examples reported in the literature. PUFs can also
be classified based on their security characteristics, specifically the
size of their challenge--response pair (CRP) space. This classification
distinguishes weak PUFs from strong PUFs, without implying a difference
in security strength. Weak PUFs support a limited number of CRPs and are
primarily used for applications such as cryptographic key generation,
device identification, and seeding pseudo-random number generators
{[}98{]}. In typical implementations, PUF responses are kept internal to
the device and are not directly exposed during normal operation.

\textbf{Table 4.} Common Attacks Against PUFs and Typical
Countermeasures

\begin{longtable}[]{@{}
  >{\raggedright\arraybackslash}p{(\columnwidth - 6\tabcolsep) * \real{0.1494}}
  >{\raggedright\arraybackslash}p{(\columnwidth - 6\tabcolsep) * \real{0.1432}}
  >{\raggedright\arraybackslash}p{(\columnwidth - 6\tabcolsep) * \real{0.3086}}
  >{\raggedright\arraybackslash}p{(\columnwidth - 6\tabcolsep) * \real{0.3988}}@{}}
\toprule()
\begin{minipage}[b]{\linewidth}\raggedright
\textbf{Attack Type}
\end{minipage} & \begin{minipage}[b]{\linewidth}\raggedright
\textbf{Targeted PUFs}
\end{minipage} & \begin{minipage}[b]{\linewidth}\raggedright
\textbf{Attack Description}
\end{minipage} & \begin{minipage}[b]{\linewidth}\raggedright
\textbf{Common Countermeasures}
\end{minipage} \\
\midrule()
\endhead
\textbf{Modeling Attacks} & Strong PUFs & Use machine learning or
statistical models to predict responses from observed CRPs & CRP
obfuscation, non-linear architecture, controlled access \\
\textbf{Physical Attacks} & All PUFs & Invasive or semi-invasive probing
of gates, delays, or memory cells & Layout obfuscation, delay-wire
shielding, tamper-evident design \\
\textbf{Side-Channel Attacks} & Weak \& Strong PUFs & Exploit power, EM,
or timing leakage during PUF evaluation or ECC & Constant-time logic,
masking, noise injection \\
\textbf{Cloning Attacks} & All PUFs & Attempt to physically replicate a
PUF's behavior & Inherent manufacturing randomness; exact physical
cloning remains infeasible with current technology \\
\bottomrule()
\end{longtable}

Table 4 provides an overview of the major classes of attacks targeting
PUF-based systems, their affected PUF types, and commonly adopted
countermeasures. Typical examples of weak PUFs include SRAM PUFs,
Ring-Oscillator PUFs, and RS latch-based PUFs. Strong PUFs, in contrast,
support a large CRP space and are commonly used in challenge--response
authentication protocols. In such systems, an external verifier may have
access to CRPs during enrollment or authentication, but it should remain
computationally infeasible for an adversary to predict valid responses
for unseen challenges within a practical time frame. Arbiter-based PUFs
and bistable ring PUFs are representative examples of strong PUF
architectures. Importantly, the distinction between weak and strong PUFs
is purely functional and relates only to CRP scalability rather than
inherent resistance to attacks.

\textbf{9- Types of PUFs Used in IoT Hardware}

Physical Unclonable Functions (PUFs) have been widely adopted in
Internet of Things (IoT) hardware to provide lightweight,
hardware-rooted security primitives such as device authentication,
identity generation, and secure key storage. Due to strict constraints
on power, cost, and computational resources in IoT devices, only a
subset of PUF architectures are practically suitable {[}99{]}. The most
commonly used PUF types in IoT platforms are summarized below.

\textbf{9-1- SRAM PUF}

SRAM PUFs exploit the random startup values of uninitialized SRAM cells
caused by manufacturing variations. When powered on, each SRAM cell
stabilizes to either `0' or `1' in a device-specific manner. SRAM PUFs
are widely used in IoT systems because SRAM is already present in most
microcontrollers, making this approach cost-effective and energy
efficient {[}100{]}. However, environmental variations such as
temperature and voltage fluctuations may affect response stability,
requiring error correction mechanisms.

\textbf{9-2- Ring Oscillator (RO) PUF}

RO-PUFs exploit frequency variations among identically designed ring
oscillators that arise from fabrication-induced process variations. In
this architecture, the PUF response is generated by comparing the
oscillation frequencies of selected oscillator pairs {[}101{]}. RO-PUFs
are widely considered suitable for IoT devices implemented on FPGAs and
ASICs due to their relatively good reliability and a moderate
challenge--response space. However, RO-PUF designs typically require
additional hardware resources, including counters and frequency
measurement circuitry, which can increase area and power consumption
{[}101{]}. A key implementation challenge in RO-PUFs is the inherent
trade-off between measurement precision and noise sensitivity during
frequency comparison. In particular, phase noise in ring oscillators can
lead to frequency measurement errors, reducing response stability and
increasing intra-device variability. Therefore, reducing oscillator
phase noise becomes critical for reliable RO-PUF operation. Design
techniques such as body biasing have been investigated to simultaneously
improve phase noise characteristics and extend the frequency tuning
range in ring VCOs {[}123{]}. Such approaches may enhance the robustness
of RO-PUF frequency comparisons under varying environmental conditions,
including temperature and voltage fluctuations.

\textbf{9-3-Arbiter PUF}

Arbiter PUFs generate responses based on delay differences between two
symmetric signal paths. A challenge determines the configuration of
multiplexers along the paths, and an arbiter decides which signal
arrives first {[}102{]}. Arbiter PUFs offer a large challenge--response
space and are often classified as strong PUFs. However, they are
vulnerable to modeling attacks using machine learning, which limits
their applicability in exposed IoT environments.

\textbf{9-4- Flash PUF}

Flash PUFs exploit threshold voltage variations in flash memory cells.
Since flash memory is commonly available in IoT devices for firmware
storage, Flash PUFs can be implemented without additional hardware. They
are primarily used for device identification rather than high-security
authentication due to limited entropy and susceptibility to aging and
retention loss.

\textbf{9-5- FPGA-Based PUF}

FPGA-based PUFs leverage configurable logic blocks, routing delays, or
embedded memory structures available on reconfigurable platforms. These
PUFs are attractive for prototyping and flexible IoT deployments
{[}103{]}. However, their security may depend on placement and routing
constraints, which can be influenced by design tools.

\textbf{9-6-Emerging PUFs (Memristor, ReRAM)}

Emerging PUF designs based on memristive devices and resistive
random-access memory (ReRAM) leverage intrinsic nanoscale resistance
variations to generate unique device fingerprints. These architectures
can provide high entropy, compact footprint, and low power consumption,
making them promising candidates for next-generation IoT security
solutions {[}104{]}. However, despite these advantages, such PUFs remain
largely limited to experimental prototypes and early-stage commercial
implementations.

\textbf{Table 5:} Common PUF Types Used in IoT Hardware

\begin{longtable}[]{@{}
  >{\raggedright\arraybackslash}p{(\columnwidth - 8\tabcolsep) * \real{0.1727}}
  >{\raggedright\arraybackslash}p{(\columnwidth - 8\tabcolsep) * \real{0.1973}}
  >{\raggedright\arraybackslash}p{(\columnwidth - 8\tabcolsep) * \real{0.1424}}
  >{\raggedright\arraybackslash}p{(\columnwidth - 8\tabcolsep) * \real{0.2185}}
  >{\raggedright\arraybackslash}p{(\columnwidth - 8\tabcolsep) * \real{0.2691}}@{}}
\toprule()
\begin{minipage}[b]{\linewidth}\raggedright
\textbf{PUF Type}
\end{minipage} & \begin{minipage}[b]{\linewidth}\raggedright
\textbf{Underlying Principle}
\end{minipage} & \begin{minipage}[b]{\linewidth}\raggedright
\textbf{IoT Suitability}
\end{minipage} & \begin{minipage}[b]{\linewidth}\raggedright
\textbf{Main Advantages}
\end{minipage} & \begin{minipage}[b]{\linewidth}\raggedright
\textbf{Limitations}
\end{minipage} \\
\midrule()
\endhead
SRAM PUF & SRAM startup state & High & No extra hardware, low power &
Sensitive to noise \\
RO-PUF & Frequency variation & Medium--High & Good stability,
reconfigurable & Area and power overhead \\
Arbiter PUF & Path delay race & Medium & Large CRP space & Vulnerable to
ML attacks under large CRP exposure \\
Flash PUF & Threshold voltage variation & Medium & Uses existing flash
memory & Aging, limited entropy \\
FPGA-based PUF & Logic/routing variation & Medium & Flexible
implementation & Tool-dependent behavior \\
Memristor / ReRAM PUF & Resistance variation & Emerging & High entropy,
low power & Immature technology \\
\bottomrule()
\end{longtable}

Table 5 summarizes the main PUF architectures commonly deployed in IoT
hardware, highlighting their operating principles, suitability for
constrained devices, and key trade-offs. The comparison demonstrates
that while SRAM and RO-PUFs dominate current IoT deployments, emerging
memory-based PUFs hold strong potential for future ultra-low-power
security applications.

\textbf{9-7- Hardware Roots of Trust}

A Hardware Root of Trust (HRoT) forms the foundational security anchor
of modern computing systems by establishing an immutable and verifiable
base upon which all higher-level security guarantees depend {[}105{]}.
In the context of IoT authentication, edge intelligence, and AI model
integrity, HRoT mechanisms are indispensable for ensuring that devices,
firmware, and deployed machine learning models remain authentic,
untampered, and trustworthy throughout their operational lifecycle.
Traditional software-based security mechanisms rely on mutable
components such as operating systems, virtual machines, or container
runtimes, which can be compromised through privilege escalation,
supply-chain attacks, or runtime exploitation {[}106{]}. In contrast,
HRoT mechanisms leverage physically protected hardware primitives that
provide security assurances independent of the software stack. These
primitives act as the first link in a chain of trust, enabling secure
boot, attestation, key protection, and integrity verification before any
untrusted code is executed {[}105,106{]}.

One of the most established HRoT implementations is the Trusted Platform
Module (TPM), standardized by the Trusted Computing Group. TPM provides
isolated execution and protected storage for cryptographic keys,
ensuring that sensitive secrets are never exposed to system memory.
Central to TPM functionality are Platform Configuration Registers
(PCRs), which store cryptographic hashes of firmware, bootloaders,
kernel images, and configuration parameters. PCRs are updated through an
irreversible extend operation, forming a cryptographically verifiable
measurement chain that reflects the system's boot and runtime state.
This PCR-based measurement chain enables remote attestation, allowing
verifiers to confirm whether a device or platform is operating in an
expected and uncompromised state. For AI-enabled IoT systems, this
capability is critical: it allows not only the verification of firmware
and operating systems but also the integrity of AI inference engines,
deployed neural network models, and security-critical libraries.
However, TPM-centric approaches face practical limitations in
large-scale, resource-constrained IoT deployments, including cost, power
consumption, and limited flexibility across heterogeneous hardware.
Physical Unclonable Functions (PUFs) provide a complementary and often
more lightweight HRoT mechanism, particularly well-suited for
constrained edge devices. PUFs exploit uncontrollable manufacturing
variations in silicon to generate device-unique and unpredictable
responses. Unlike traditional keys stored in non-volatile memory,
PUF-derived secrets are never permanently stored, significantly reducing
the attack surface against physical extraction, cloning, and invasive
attacks.

From an authentication perspective, PUFs enable device-intrinsic
identity, allowing each IoT node to be uniquely authenticated without
requiring centralized per-device key provisioning or dedicated secure
key storage. More importantly, PUFs can act as roots of trust for
cryptographic key generation, binding security credentials directly to
physical hardware properties. This capability is especially valuable in
distributed IoT and edge AI ecosystems where devices operate
autonomously and intermittently connect to the cloud (see Figure 6).

\includegraphics[width=4.849in,height=3.55779in]{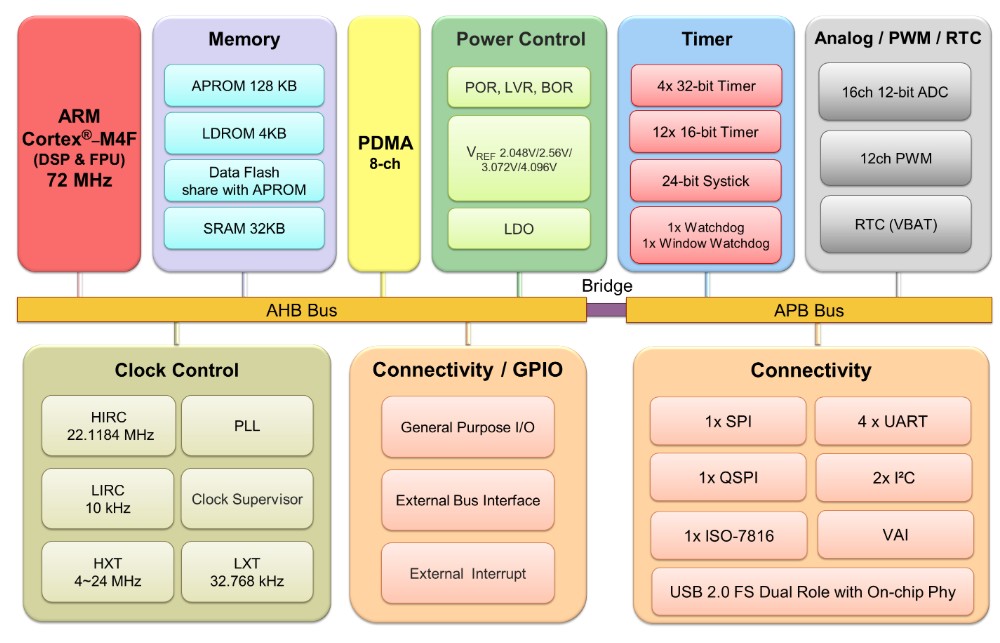}

\textbf{Figure 7.} Typical microcontroller architecture used in edge IoT devices; hardware root-of-trust components such as secure boot, PUF, or TPM are not explicitly shown\textbf{.}

Recent research has extended HRoT concepts into containerized and
virtualized environments, where traditional hardware trust anchors are
not directly accessible to lightweight workloads. By combining TPM PCR
measurements with container-specific runtime features and encapsulating
the resulting trust material within Trusted Execution Environments
(TEEs), it becomes possible to construct container-oriented PUFs
(CPUFs). These hybrid mechanisms preserve hardware-rooted trust while
supporting modern DevOps, microservices, and edge AI deployment models.
Figure 7 illustrates a hardware-enforced chain of trust built around 
eFuse-based roots of trust, secure programming paths, and 
access-controlled instruction and data memories within a RISC-V processing core. 
Security-critical components, including non-volatile memory, access control logic, 
and secure communication buses, establish device integrity from the boot stage. 
Meanwhile, trusted data paths support on-device signal processing and AI inference 
(e.g., FFT and SVM). This architecture serves as a representative example of how hardware roots of trust can ensure device authentication, firmware integrity, and secure AI model execution 
in resource-constrained edge environments. {[}107{]}.

Within AI-driven IoT systems, Hardware Roots of Trust (HRoT) mechanisms
play a crucial role in AI model integrity protection. Edge devices
increasingly execute lightweight machine learning inference locally to
reduce latency and bandwidth consumption. For example, IoT camera
systems for wildlife monitoring perform on-device deep learning analysis
before transmitting processed results to cloud servers {[}124{]}. While
such edge-AI deployment improves responsiveness and scalability, it also
introduces new attack surfaces, including model tampering, unauthorized
replacement, poisoning, rollback attacks, and intellectual property
extraction. By binding AI model hashes or feature fingerprints to TPM
Platform Configuration Registers (PCRs) or PUF-derived keys, systems can
ensure that inference engines operate only when both the hardware state
and model integrity meet predefined trust conditions {[}107{]}. This
hardware--AI binding significantly raises the bar for adversaries
attempting to manipulate AI behavior or exfiltrate intellectual
property.

In national security and critical infrastructure contexts, HRoT
mechanisms also mitigate hardware interdiction and supply-chain threats,
where adversaries attempt to insert malicious logic before deployment.
Since PUF behavior cannot be cloned or predicted---even by the
manufacturer---PUF-based HRoT architectures provide strong resistance
against counterfeit hardware and unauthorized replication. In summary,
Hardware Roots of Trust serve as the cornerstone of secure IoT
authentication and AI model integrity. TPMs offer standardized,
measurement-based trust for complex platforms, while PUFs provide
lightweight, scalable, and physically grounded security for constrained
and distributed environments {[}108{]}. Their integration, especially in
edge and containerized systems, enables resilient, hardware-backed trust
architectures capable of supporting next-generation IoT and AI
deployments.

\textbf{Table 6.} Comparison of Trust Anchor Mechanisms for Secure AI
Model Integrity and Deployment in IoT and Edge Systems

\begin{longtable}[]{@{}
  >{\raggedright\arraybackslash}p{(\columnwidth - 12\tabcolsep) * \real{0.1436}}
  >{\raggedright\arraybackslash}p{(\columnwidth - 12\tabcolsep) * \real{0.1265}}
  >{\raggedright\arraybackslash}p{(\columnwidth - 12\tabcolsep) * \real{0.1267}}
  >{\raggedright\arraybackslash}p{(\columnwidth - 12\tabcolsep) * \real{0.1297}}
  >{\raggedright\arraybackslash}p{(\columnwidth - 12\tabcolsep) * \real{0.1364}}
  >{\raggedright\arraybackslash}p{(\columnwidth - 12\tabcolsep) * \real{0.1702}}
  >{\raggedright\arraybackslash}p{(\columnwidth - 12\tabcolsep) * \real{0.1668}}@{}}
\toprule()
\begin{minipage}[b]{\linewidth}\raggedright
\textbf{Method}
\end{minipage} & \begin{minipage}[b]{\linewidth}\raggedright
\textbf{Trust Anchor}
\end{minipage} & \begin{minipage}[b]{\linewidth}\raggedright
\textbf{Key Storage}
\end{minipage} & \begin{minipage}[b]{\linewidth}\raggedright
\textbf{Size}
\end{minipage} & \begin{minipage}[b]{\linewidth}\raggedright
\textbf{IoT Suitability}
\end{minipage} & \begin{minipage}[b]{\linewidth}\raggedright
\textbf{AI Model Integrity Support}
\end{minipage} & \begin{minipage}[b]{\linewidth}\raggedright
\textbf{Main Limitation}
\end{minipage} \\
\midrule()
\endhead
TPM + PCR & Dedicated security chip & Secure non-volatile memory &
Medium & Limited for ultra-constrained devices & Strong (via measurement
\& attestation) & Cost, power, integration overhead \\
Silicon PUF & Physical process variations & Not stored (generated on
demand) & High & Excellent & Moderate--Strong (key binding,
authentication) & Environmental sensitivity \\
FPGA RO-PUF & Delay-based oscillator variations & Not stored &
Medium--High & Good (FPGA-based edge nodes) & Strong (ML-resistant
identity) & FPGA resource usage \\
CPUF (TPM + Container Features) & Hybrid hardware--software &
TEE-protected & High & Good for edge/cloud & Strong (runtime + model
binding) & Complexity, orchestration overhead \\
Software-only Trust & OS / VM & Memory or disk & High & Poor & Weak &
Easily bypassed or tampered \\
\bottomrule()
\end{longtable}

Table 6 compares representative trust anchor mechanisms for securing AI
model integrity and deployment in IoT and edge systems, highlighting the
trade-offs between security strength, scalability, and implementation
overhead. Hardware-rooted solutions such as TPMs and silicon-based PUFs
provide strong guarantees for device identity, secure boot, and
integrity verification by anchoring trust in immutable physical
properties. TPM-based approaches leverage measurement and attestation
through PCRs to protect firmware and AI workloads, but their cost, power
consumption, and integration complexity limit applicability in
ultra-constrained devices. In contrast, PUF-based roots of trust
generate secrets on demand without persistent storage, offering
excellent scalability and suitability for resource-limited IoT nodes,
albeit with sensitivity to environmental variations. FPGA-based RO-PUFs
occupy an intermediate design space, combining reconfigurability with
hardware-rooted identity, making them particularly attractive for edge
nodes requiring adaptable security features. Hybrid mechanisms such as
container-oriented PUFs (CPUFs) extend hardware trust into virtualized
and containerized environments by combining TPM measurements with
runtime protections, enabling strong AI model binding and attestation
across edge--cloud deployments at the cost of increased system
complexity. Software-only trust mechanisms, while highly scalable and
flexible, lack strong tamper resistance and are vulnerable to bypass and
runtime manipulation. Overall, the comparison underscores that
hardware-anchored trust mechanisms are essential for protecting AI model
integrity and device authenticity in IoT systems, while hybrid
approaches provide a promising path toward secure and scalable edge
intelligence.

\textbf{10-PUF Architectures for IoT Hardware}
Physical Unclonable Functions (PUFs) have been widely recognized as
effective hardware security primitives for IoT systems due to their
ability to generate device-specific identities based on intrinsic
physical characteristics. Unlike traditional authentication mechanisms
that rely on stored credentials, PUFs exploit unavoidable manufacturing
variations that arise during integrated circuit fabrication {[}109{]}.
These variations, which include differences in threshold voltage,
transistor behavior, capacitance, and other physical parameters, cannot
be precisely controlled or replicated, even by the original
manufacturer. As a result, each device exhibits unique
challenge--response behavior that can be used for authentication,
identification, and secret generation {[}109{]}. Table 7 presents a
structured classification of security attacks mapped to the main layers
of the IoT architecture, namely the encryption (or security) layer,
perception layer, network layer, and application layer. This taxonomy
highlights how different attack vectors target distinct functional
components of IoT systems, reflecting the heterogeneous and
multi-layered nature of IoT security threats.

\textbf{Table 7.} Layer-wise taxonomy of security attacks in the IoT
architecture

\begin{longtable}[]{@{}
  >{\raggedright\arraybackslash}p{(\columnwidth - 6\tabcolsep) * \real{0.3283}}
  >{\raggedright\arraybackslash}p{(\columnwidth - 6\tabcolsep) * \real{0.2185}}
  >{\raggedright\arraybackslash}p{(\columnwidth - 6\tabcolsep) * \real{0.2246}}
  >{\raggedright\arraybackslash}p{(\columnwidth - 6\tabcolsep) * \real{0.2286}}@{}}
\toprule()
\begin{minipage}[b]{\linewidth}\raggedright
\textbf{Cryptographic / Encryption-layer attacks}
\end{minipage} & \begin{minipage}[b]{\linewidth}\raggedright
\textbf{Perception-layer attacks}
\end{minipage} & \begin{minipage}[b]{\linewidth}\raggedright
\textbf{Network-layer attacks}
\end{minipage} & \begin{minipage}[b]{\linewidth}\raggedright
\textbf{Application-layer attacks}
\end{minipage} \\
\midrule()
\endhead
Side-channel attack & Device (node) tampering & Sybil attack & Viruses
and worms \\
Radio-frequency (RF) interference & Physical damage & Route manipulation
attack & Spyware and adware \\
Man-in-the-middle (MITM) attack & Node jamming & Sinkhole attack &
Trojan horse \\
Malicious node insertion & RFID spoofing & Denial-of-service (DoS) &
Malicious scripts \\
Cryptanalytic attacks & RFID cloning & MITM attack & Ransomware \\
\bottomrule()
\end{longtable}

At the encryption layer, attacks primarily aim to compromise
cryptographic mechanisms and secret information. Side-channel attacks
exploit physical leakages such as power consumption or timing
variations, while man-in-the-middle and replay attacks target weaknesses
in key exchange and authentication protocols. Cryptanalytic and
sleep-deprivation attacks further threaten the confidentiality and
availability of cryptographic operations by exhausting device resources
or breaking encryption schemes {[}110{]}. In IoT environments, where
devices are often resource-constrained and deployed in untrusted or
physically exposed locations, storing cryptographic keys in non-volatile
memory introduces significant security risks. PUF-based techniques
address this challenge by deriving secrets dynamically from hardware
properties, thereby reducing the attack surface associated with key
storage. This capability makes PUFs particularly attractive for
large-scale IoT deployments, where device authentication is critical for
maintaining secure and trustworthy communication among heterogeneous
nodes.

\textbf{11- PUFs and AI-Enabled Hardware Security}

Artificial intelligence (AI), machine learning (ML), and deep learning
(DL) have become increasingly influential in hardware security,
including the design and analysis of PUF-based systems. The rapid
advancement of AI-driven technologies has led to a substantial increase
in the volume of data processed by modern computing platforms and has
enabled intelligent applications across diverse domains. In the context
of PUFs, AI techniques are explored both as defensive tools to enhance
security and as offensive tools that expose vulnerabilities in existing
designs {[}111{]}. On the defensive side, researchers have investigated
AI-enabled hardware security detection techniques capable of identifying
malicious activity in real time. Machine learning--based detectors can
be embedded directly within microprocessor hardware and integrated with
high-performance computing data paths.

\includegraphics[width=4.17147in,height=3.6905in]{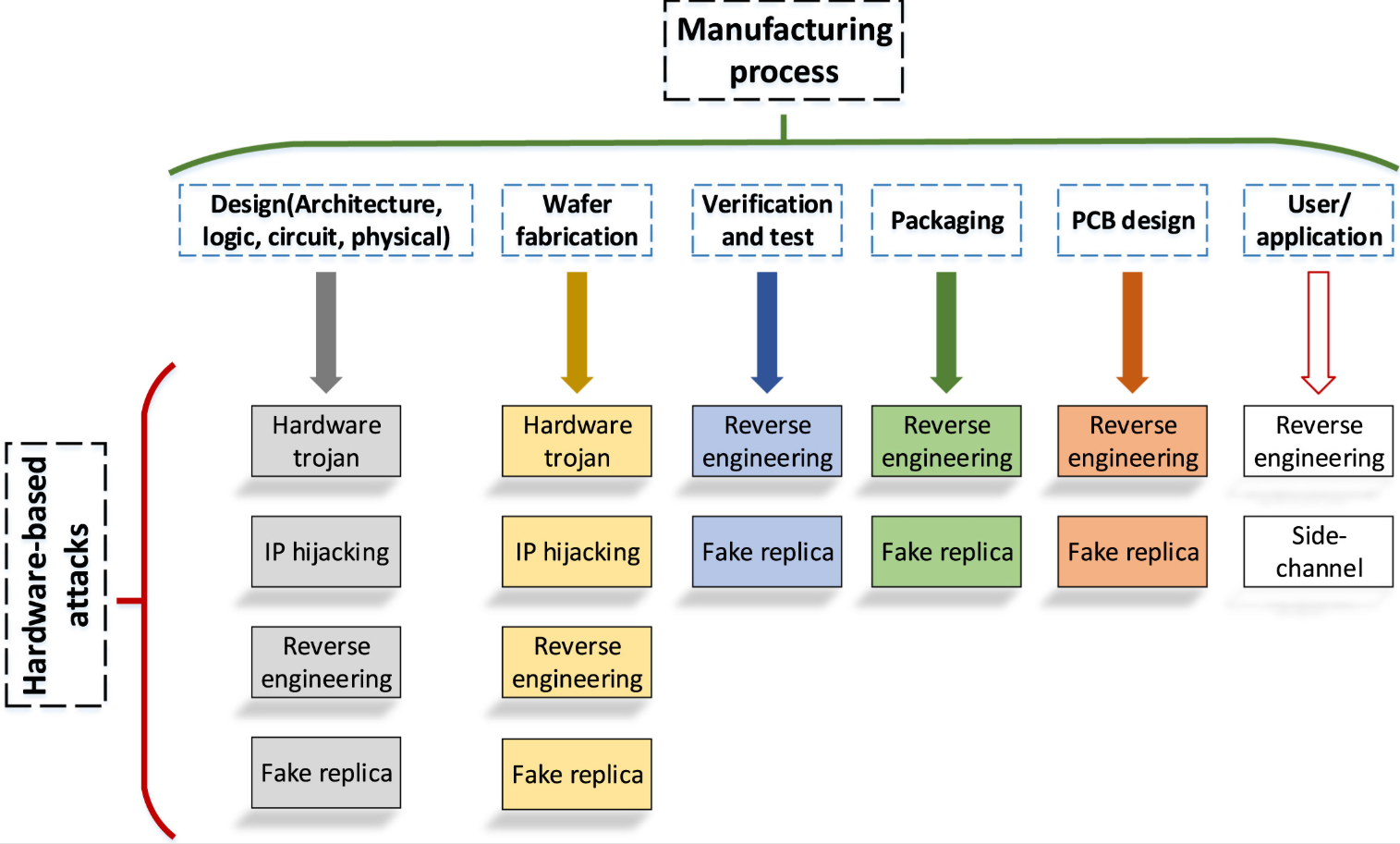}

\textbf{Figure 8.} Hardware-based attack vectors across the integrated
circuit manufacturing and deployment lifecycle

Compared with software-based approaches, hardware-level detection
mechanisms offer lower overhead and faster response times, often
operating within a few clock cycles. These characteristics enable rapid
identification of malicious code patterns and support the deployment of
secure computing systems at the end host. Figure 8 illustrates how
hardware-based attacks can be introduced at different stages of the
integrated circuit (IC) lifecycle, spanning from initial design to
end-user deployment. The top portion of the diagram outlines the
standard manufacturing and deployment pipeline, including design
(architecture, logic, circuit, and physical layout), wafer fabrication,
verification and testing, packaging, PCB design, and final user
application. Each stage represents a potential attack surface where
adversaries may exploit supply-chain complexity and limited trust
assumptions.

At the same time, literature highlights ongoing challenges associated
with AI-driven modeling attacks on PUFs. Adversaries can collect
challenge--response pairs and use machine learning algorithms to
construct surrogate models that approximate PUF behavior {[}112{]}. Such
attacks undermine the unpredictability and uniqueness upon which
PUF-based security relies. As AI and ML techniques continue to evolve,
attackers gain improved capability to approximate increasingly complex
PUF architectures, reinforcing the need for designs that can withstand
learning-based attacks {[}112{]}.

\textbf{11-1- Early Development of PUFs and Architectural Evolution}

The conceptual foundations of Physical Unclonable Functions (PUFs) can
be linked to early physical identification mechanisms such as biometric
fingerprint recognition which introduced the idea of leveraging
intrinsic physical characteristics for security purposes. In the
hardware security domain, these ideas evolved through the development of
physical one-way functions and physical random functions, ultimately
leading to what are now formally known as PUFs. Modern PUF architectures
exploit inherent and uncontrollable physical variations introduced
during semiconductor manufacturing to generate device-unique
challenge--response behavior that cannot be precisely replicated, even
by the original manufacturer. PUFs are commonly categorized
according to their physical implementation and the size of their
challenge--response pair (CRP) space {[}113{]}. Weak PUFs generate a
limited number of stable responses and are typically employed for device
identification, key derivation, or secure storage applications, whereas
strong PUFs support a large CRP space and are more suitable for
challenge--response authentication protocols. These architectural
distinctions directly influence how PUFs are deployed in IoT systems and
determine their susceptibility to modeling attacks, environmental
variation, and long-term reliability concerns {[}114{]}.

Despite significant research efforts aimed at improving PUF robustness
and security, open challenges remain regarding their resilience against
emerging hardware-level threats, including sophisticated physical
attacks, invasive probing, and learning-assisted adversaries. As IoT
systems continue to scale and operate in increasingly adversarial
environments, these concerns motivate ongoing research into
modeling-resistant architectures, controlled CRP exposure, and hybrid
hardware--software protection mechanisms. Recently, PUF
architectures have increasingly focused on improving resistance to
modern attack vectors, particularly those enabled by machine
learning--based modeling. Machine learning--resistant PUF designs seek
to preserve the intrinsic security benefits of hardware variability
while mitigating vulnerabilities exposed by modeling attacks using
algorithms such as support vector machines, logistic regression,
artificial neural networks, and evolutionary optimization techniques
{[}115{]}. These approaches aim to reduce the feasibility of
constructing accurate predictive models of PUF behavior; however, the
added architectural complexity often introduces new challenges related
to area overhead, power consumption, reliability, and implementation
robustness.

Memristive PUFs represent another active direction in architectural
innovation {[}116{]}. By exploiting the intrinsic resistance variability
and nonlinear switching behavior of memristive devices, these PUFs can
achieve high entropy and increased modeling complexity. Some designs
further enhance security by enabling PUF regeneration rather than
relying on permanently stored secrets, thereby reducing the attack
surface associated with non-volatile key storage. While the nonlinear
behavior of memristive devices can significantly increase attack
complexity and modeling time, practical concerns remain regarding
stability, endurance, and susceptibility to environmental variation.
Device-level nonlinearity and instability, as reported in related
memristive systems, may negatively impact long-term reliability and
repeatability, indicating that modeling attacks and robustness
challenges remain active research issues for memristive PUF
implementations {[}117{]}. Erasable PUFs introduce the capability to
selectively invalidate individual challenge--response pairs (CRPs)
without affecting the remaining PUF behavior, with the goal of limiting
adversarial learning through repeated CRP observation. By permanently
removing CRPs after use, erasable PUFs reduce the amount of data
available to attackers and are therefore particularly effective in
mitigating machine learning--based modeling attacks and large-scale CRP
harvesting. However, erasable PUFs do not inherently protect against
side-channel leakage, which arises from power, electromagnetic, or
timing information during PUF evaluation and must be addressed through
complementary circuit-level countermeasures. Despite their
advantages, erasable PUFs can be compromised if adversaries gain
unauthorized access to erase control mechanisms or associated circuitry,
potentially enabling response manipulation, forced invalidation, or
denial-of-service attacks. Consequently, ongoing research continues to
explore architectural enhancements that combine controlled CRP exposure,
learning resistance, and secure erase mechanisms while maintaining
reliability under environmental and operational variations.

\textbf{Table 8.} Evolution of PUF architectures and associated
limitations

\begin{longtable}[]{@{}
  >{\raggedright\arraybackslash}p{(\columnwidth - 8\tabcolsep) * \real{0.1807}}
  >{\raggedright\arraybackslash}p{(\columnwidth - 8\tabcolsep) * \real{0.2307}}
  >{\raggedright\arraybackslash}p{(\columnwidth - 8\tabcolsep) * \real{0.1877}}
  >{\raggedright\arraybackslash}p{(\columnwidth - 8\tabcolsep) * \real{0.1932}}
  >{\raggedright\arraybackslash}p{(\columnwidth - 8\tabcolsep) * \real{0.2077}}@{}}
\toprule()
\begin{minipage}[b]{\linewidth}\raggedright
\textbf{PUF category / architecture}
\end{minipage} & \begin{minipage}[b]{\linewidth}\raggedright
\textbf{Key characteristics}
\end{minipage} & \begin{minipage}[b]{\linewidth}\raggedright
\textbf{Typical use cases}
\end{minipage} & \begin{minipage}[b]{\linewidth}\raggedright
\textbf{Main security advantage}
\end{minipage} & \begin{minipage}[b]{\linewidth}\raggedright
\textbf{Key limitations}
\end{minipage} \\
\midrule()
\endhead
Early physical identification mechanisms & Based on intrinsic physical
characteristics (e.g., fingerprint-inspired concepts) & Physical
identification & Introduced the idea of uniqueness from physical
properties & Not directly applicable to modern IC-based security \\
Physical one-way and random functions & Intermediate concepts leading to
PUFs & Hardware security primitives & One-way behavior and randomness &
Limited formalization and scalability \\
Conventional PUFs & Exploit manufacturing variations to generate CRPs &
Device identification and authentication & Unique, device-specific
responses & Sensitivity to environmental variations \\
Weak PUFs & Limited number of CRPs & Identification, secure key storage
& Low complexity, stable responses & Not suitable for large-scale
authentication \\
Strong PUFs & Large CRP space & Authentication protocols & Increased
resistance to brute-force attacks & Greater exposure to modeling
attacks \\
ML-resistant PUFs & Designed to resist learning-based modeling & Secure
authentication under ML threat & Reduced predictability of CRPs &
Increased architectural complexity \\
Symmetric function--based memristive PUFs & Use symmetric functions and
memristor nonlinearity & Hardware-based key generation & Avoid explicit
storage of secrets & Reliability and susceptibility concerns; modeling
attacks remain \\
Erasable PUFs & Support selective removal of CRPs & Limiting adversarial
learning & Reduced CRP exposure over time & Vulnerable if erase/control
mechanisms are compromised \\
\bottomrule()
\end{longtable}

Table 8 summarizes the evolution of Physical Unclonable Function (PUF)
architectures and highlights the main security motivations and
limitations associated with each stage of development. The table begins
with early physical identification mechanisms, which introduced the
fundamental idea of deriving security from intrinsic physical
characteristics. Although these approaches inspired later developments,
they were not directly suited for integrated circuit--level security and
therefore served mainly as conceptual foundations.

\textbf{12-Threats of Counterfeit ICs and Mitigation Strategies}

\textbf{12-1- Counterfeit ICs as a Supply-Chain Threat in IoT}

The widespread outsourcing of semiconductor fabrication and packaging to
offshore manufacturing facilities has introduced multiple vulnerability
paths within the integrated circuit (IC) supply chain. In this
globalized production model, detecting counterfeit or maliciously
altered components remains a significant challenge, which is further
exacerbated as ICs are increasingly embedded in large-scale Internet of
Things (IoT) deployments. Since IoT devices are often produced in high
volume and deployed in environments with limited physical and runtime
security monitoring, compromises at the hardware level can propagate
across entire systems and lead to severe safety, reliability, and
security consequences. Beyond technical risks, counterfeiting and
piracy impose substantial economic and societal costs. Various industry
and policy reports estimate that global losses attributed to
counterfeiting and piracy amount to multiple trillions of U.S. dollars
annually, with reported figures ranging from hundreds of billions to
several trillion dollars worldwide, depending on scope and methodology
{[}118{]}. Counterfeiting is also estimated to cost G20 economies on the
order of hundreds of billions of dollars per year. In addition to direct
financial losses, counterfeit ICs undermine system safety, erode trust
in critical infrastructure, and contribute to long-term trade imbalances
and security vulnerabilities in global supply chains {[}118{]}.

\textbf{12-2-Deception-Based Authentication Against Fake Replica Attacks}

A deception-driven authentication protocol has been proposed as a
preventive defense strategy against fake replica and modeling-based
attacks. The main idea is to mislead adversaries by exposing a training
set dominated by hoax responses, which disrupts the learning process of
machine learning--based attackers {[}119{]}. Through adversarial machine
learning techniques, errors are intentionally introduced into the
attacker's surrogate model by transmitting ``poisoned''
challenge--response pairs (CRPs). By reducing the predictive accuracy of
the adversary's model, this approach makes it harder to replicate PUF
behavior or predict responses to previously unseen challenges. Deception
is therefore positioned as a strategy that delays, confuses, and
disrupts attacker workflows. Since adversaries must spend effort trying
to separate genuine signals from deceptive ones, this defensive method
is expected to impose economic cost on attackers and discourage them
from continuing the attack, thereby weakening attempts to build
effective PUF clones.

\textbf{12-3- Machine Learning for Detecting Counterfeit ICs and Hardware Trojans}

IC counterfeiting and PUF-based hardware trojans are increasingly viewed
as serious threats to the semiconductor ecosystem because attackers can
embed malicious logic directly into hardware. Conventional inspection
and verification methods often fail to reliably identify these
sophisticated manipulations. To address this gap, machine learning
techniques are increasingly used to automate and improve the inspection
process. In particular, parametric data collected from on-chip sensors
can be analyzed and categorized using support vector machines (SVMs),
which have been reported as effective for identifying recycled ICs and
detecting hardware trojans in real time. In addition to SVMs, other
learning methods such as random forest and multilayer perceptron (MLP)
have been applied in defense contexts, including protection against
microarchitectural side-channel attacks. These efforts reflect a broader
movement toward ML-assisted security evaluation, where data-driven
classification supports faster and more scalable detection than manual
inspection alone.

\textbf{12-4-Architectural Enhancements for Side-Channel Resistance}
Beyond detection and post-deployment monitoring, architectural-level
security enhancements have been explored to proactively reduce
side-channel leakage at its physical source. Rather than treating
side-channel attacks solely as a software or runtime problem, these
approaches integrate security mechanisms directly into circuit and
system architectures to limit information leakage through power, timing,
or electromagnetic channels {[}120{]}. One research direction
exploits the aging behavior and nonlinear characteristics of emerging
devices such as memristors to support secure neuromorphic and adaptive
computing systems. The intrinsic variability and temporal evolution of
memristive elements can introduce uncertainty that complicates
side-channel analysis, while simultaneously enabling functionality such
as in-memory computation and low-power inference. Another line of work
introduces machine learning--assisted power compensation and balancing
circuits, where learned models dynamically adjust power profiles to mask
data-dependent variations. Compared to traditional constant-power or
noise-injection techniques, these ML-driven approaches aim to achieve
improved side-channel resistance with reduced power and area overhead.
Together, these developments highlight that machine learning--based
techniques are not limited to software-level anomaly detection or
post-processing analysis. Instead, ML can be embedded within hardware
architectures themselves to enable adaptive, low-overhead mitigation of
side-channel leakage. Such architectural enhancements complement
PUF-based authentication and hardware-rooted trust mechanisms by
strengthening resistance against side-channel exploitation, hardware
Trojans, and counterfeiting across the IoT device lifecycle {[}121{]}.

\textbf{13- conclusion}

Recent advances in Physical Unclonable Function (PUF) architectures have
produced a wide spectrum of approaches aimed at strengthening the
security of Internet of Things (IoT) devices, particularly in the
presence of counterfeit hardware, replica attacks, and learning-based
modeling threats. While significant progress has been achieved in
improving entropy sources, architectural complexity, and attack
resistance, meaningful comparison across existing solutions remains
challenging. Prior studies often rely on heterogeneous threat models,
ad-hoc evaluation methodologies, and inconsistent performance metrics,
limiting the ability to draw definitive conclusions regarding security
strength, reliability trade-offs, and deployment feasibility.

At the same time, hardware Trojan threats and supply-chain
vulnerabilities continue to expand alongside the globalization and
increasing complexity of integrated circuit (IC) manufacturing. The
distributed, multi-stage nature of modern semiconductor production
introduces multiple trust boundaries, enabling adversaries to inject
malicious modifications during design, fabrication, testing, or
recycling phases. These structural vulnerabilities are further
exacerbated by system-on-chip complexity, aggressive time-to-market
pressures, and reliance on third-party IP integration.

In parallel, IoT devices are frequently deployed in physically exposed
and resource-constrained environments, where direct hardware
manipulation, cloning, denial-of-service attacks, and learning-assisted
modeling attacks are realistic adversarial scenarios. These deployment
realities highlight the importance of developing PUF architectures that
balance robustness against machine learning--based modeling, controlled
challenge--response pair (CRP) exposure, and environmental reliability
constraints.

While traditional security solutions rely on cryptographic primitives
combined with secure key storage and trusted execution mechanisms,
PUF-based approaches provide a complementary hardware-rooted trust model
that derives secrets from intrinsic physical variations, thereby
reducing reliance on permanently stored secret keys and enabling
device-intrinsic identity. However, PUF-based systems are not immune to
architectural, modeling, and side-channel vulnerabilities, and their
practical security depends strongly on evaluation methodology and
deployment context.

Based on the surveyed literature, several research gaps remain open:\\
(i) the absence of standardized benchmarking and threat modeling
frameworks for evaluating PUF security and reliability;\\
(ii) limited robustness against increasingly sophisticated machine
learning--based modeling attacks;\\
(iii) insufficient integration of PUF-based trust anchors with AI
workload integrity verification mechanisms; and\\
(iv) unresolved challenges in supply-chain--aware Trojan detection and
lifecycle security assurance.

Addressing these gaps is essential for enabling scalable, trustworthy,
and hardware-rooted security architectures capable of supporting
next-generation AI-enabled IoT systems.

\textbf{References}

{[}1{]} Cheikh, I., Roy, S., Sabir, E., \& Aouami, R. (2026). Energy,
scalability, data and security in massive IoT: Current landscape and
future directions.~\emph{IEEE Internet of Things Journal}.

{[}2{]} Abdi, H., \& Nozari, H. (2026). Energy challenges in
transformative technologies-based super-smart city implementation.
In~\emph{Energy-Efficient Transformative Technologies for Data-Driven
Smart Cities}~(pp. 71-90). Elsevier.

{[}3{]} Anjum, M., Khan, M. A., \& Jung, H. (2026). Designing an
end-to-end sustainable IoT network: a comprehensive guideline.
In~\emph{Design and Analysis of Green and Sustainable IoT Technologies
for Future Wireless Communications}~(pp. 17-52). Academic Press.

{[}4{]} Banciu, C., \& Florea, A. (2026). AIoT at the Frontline of
Climate Change Management: Enabling Resilient, Adaptive, and Sustainable
Smart Cities.~\emph{Climate},~\emph{14}(1), 19.

{[}5{]} Maralapalle, V., Muktinutalapati, J., Chandra, B., Narala, G.
R., \& Iyer, R. (2026). Analyzing the Role of Geospatial Technologies
and Al in Urban Infrastructure Planning and the Development of Smart
Cities, Including Transportation Systems, Utilities, and Public
Services. In~\emph{Advanced Geospatial Intelligence and AI for
Environmental Resilience and Sustainable Development}~(pp. 69-100).
Cham: Springer Nature Switzerland.

{[}6{]} Goswami, S. S., \& Mondal, S. (2024). The role of 5G in
enhancing IOT connectivity: A systematic review on applications,
challenges, and future prospects. Big data and computing visions, 4(4),
314-325.

{[}7{]} Patil, R. S., \& Moantri, S. (2026). Challenges and
Opportunities in Realâ€Time Data Processing: Advancements and Limitations
in Realâ€Time Data Analytics.~\emph{Artificial Intelligence and Machine
Learning in Neurology},~\emph{2}, 647-683.

{[}8{]} Dao, T., Nguyen, M., Do, S., \& Tran, H. (2026). Cyberscurity
Threats and Defense Mechanisms in IoT network.~\emph{arXiv preprint
arXiv:2601.00556}.

{[}9{]} Wen, S. F., \& Sharma, A. (2026). Structuring Trust: A
Quantitative and Traceable Framework for Hardware Security Assurance.

{[}10{]} De Meulemeester, J., Oswald, D., Verbauwhede, I., \& Van Bulck,
J. (2026, May). Battering RAM: Low-Cost Interposer Attacks on
Confidential Computing via Dynamic Memory Aliasing. In~\emph{47th IEEE
Symposium on Security and Privacy (S\&P)}.

{[}11{]} Tashdid, I., Farheen, T., \& Rahman, S. (2026). InterPUF:
Distributed Authentication via Physically Unclonable Functions and
Multi-party Computation for Reconfigurable Interposers.~\emph{arXiv
preprint arXiv:2601.11368}.

{[}12{]} Colombier, B., \& Bossuet, L. (2014). Survey of hardware
protection of design data for integrated circuits and intellectual
properties. IET Computers \& Digital Techniques, 8(6), 274-287.

{[}13{]} JÃ¸rgensen, B. N., \& Ma, Z. G. (2026). Cybersecurity and
Resilience of Smart Grids: A Review of Threat Landscape, Incidents, and
Emerging Solutions.~\emph{Applied Sciences},~\emph{16}(2), 981.

{[}14{]} Leo, M., Tan, F., Miao, T., \& Anand, G. (2026). From threat to
trust: assessing security risks of agentic AI systems: M. Leo et
al.~\emph{International Journal of Information Security},~\emph{25}(1),
23.

{[}15{]} Shafiq, M., Gu, Z., Cheikhrouhou, O., Alhakami, W., \& Hamam,
H. (2022). The Rise of ``Internet of Things'': Review and Open Research
Issues Related to Detection and Prevention of IoTâ€Based Security
Attacks. Wireless Communications and Mobile Computing, 2022(1), 8669348.

{[}16{]} Komala, C. R., Basha, M. M., Farook, S., Niranchana, R.,
Rajendiran, M., \& Subhi, B. (2024). Smart Energy Systems-Integrated
Machine Learning, IoT, and AI Tools. In~\emph{Reshaping Environmental
Science Through Machine Learning and IoT}~(pp. 201-229). IGI Global
Scientific Publishing.

{[}17{]} Agupugo, C. P., Tochukwu, M. F. C., Ogunmoye, K. A., Mosha, A.
S., \& Sabbih, F. (2025). Review of Smart Microgrid Platform Integrating
AI and Deep Reinforcement Learning for Sustainable Energy Management.

{[}18{]} Tashdid, I., Farheen, T., \& Rahman, S. (2025, June). Safe-sip:
Secure authentication framework for system-in-package using multi-party
computation. In Proceedings of the Great Lakes Symposium on VLSI 2025
(pp. 391-396).

{[}19{]} Najafi, F., Kaveh, M., Mosavi, M. R., Brighente, A., \& Conti,
M. (2024). EPUF: An Entropy-Derived Latency-Based DRAM Physical
Unclonable Function for Lightweight Authentication in Internet of
Things.~\emph{IEEE Transactions on Mobile Computing}.

{[}20{]} Mishra, J., \& Sahay, S. K. (2025). Modern hardware security: A
review of attacks and countermeasures.~\emph{arXiv preprint
arXiv:2501.04394}.

{[}21{]} Chatterjee, D., Maitra, S., Mishra, N., Shukla, S., \&
Mukhopadhyay, D. (2025). Hardware security in the connected
world.~\emph{Wiley Interdisciplinary Reviews: Data Mining and Knowledge
Discovery},~\emph{15}(3), e70034.

{[}22{]} Robyns, P., Di Martino, M., Giese, D., Lamotte, W., Quax, P.,
\& Noubir, G. (2020, July). Practical operation extraction from
electromagnetic leakage for side-channel analysis and reverse
engineering. In~\emph{Proceedings of the 13th ACM Conference on Security
and Privacy in Wireless and Mobile Networks}~(pp. 161-172).

{[}23{]} Shwartz, O., Mathov, Y., Bohadana, M., Elovici, Y., \& Oren, Y.
(2018). Reverse engineering IoT devices: Effective techniques and
methods. IEEE Internet of Things Journal, 5(6), 4965-4976.

{[}24{]} Boubakri, M., \& Zouari, B. (2025). A Survey of RISC-V Secure
Enclaves and Trusted Execution
Environments.~\emph{Electronics},~\emph{14}(21), 4171.

{[}25{]} Rohini, C., Negi, B. S., Sood, S., Pandey, A. K., Sahu, P. K.,
\& Nagappan, B. Context-Aware Energy Management Systems for Optimizing
Power Consumption in Smart Ubiquitous Environments.

{[}26{]} Emehin, O., Akanbi, I., Emeteveke, I., \& Adeyeye, O. J.
(2024). Enhancing cybersecurity with safe and reliable AI: mitigating
threats while ensuring privacy protection.~\emph{International Journal
of Computer Applications Technology and Research, doi},~\emph{10}.

{[}27{]} Golda, A., Mekonen, K., Pandey, A., Singh, A., Hassija, V.,
Chamola, V., \& Sikdar, B. (2024). Privacy and security concerns in
generative AI: A comprehensive survey. Ieee Access, 12, 48126-48144.

{[}28{]} Johnson, R. (2025). Designing secure and scalable IoT systems:
Definitive reference for developers and engineers. HiTeX Press.

{[}29{]} Cha, W., Lee, H. J., Kook, S., Kim, K., \& Won, D. (2025). A
Lightweight Authentication and Key Distribution Protocol for XR Glasses
Using PUF and Cloud-Assisted ECC.~\emph{Sensors (Basel,
Switzerland)},~\emph{26}(1), 217.

{[}30{]} Arul Selvan, M. (2025). Utilization of Secure Bootloaders in
Embedded Systems for Ensuring Device Integrity and Preventing Firmware
Tampering Through Cryptographic Validation Mechanisms.

{[}31{]} Siyal, F., Guzzo, A., Alkhabbas, F., Sacca, D., \& Fortino, G.
(2026). Secure Supply Chain Provenance via PUF-Anchored NFTs and 6G Edge
Networks.~\emph{IEEE Wireless Communications}.

{[}32{]} Kasimatis, D., Politis, I., Pitropakis, N., Papadopoulos, P.,
\& Buchanan, W. J. (2025). Decentralised Device Identity: PUFâ€‘Driven
Soulbound Token Verification for IoT Supply Chain Security.~\emph{IEEE
Transactions on Consumer Electronics}.

{[}33{]} Tran, S., Ngo, C. T., \& Hong, J. P. (2025). A lightweight
ECC-compatible end-to-end security protocol using CRP-PUF and TRNG for
IoT devices.~\emph{IEEE Internet of Things Journal}.

{[}34{]} Vathsala, A. V., Kalyani, D., Adudhodla, M., Saraf, S.,
Manellore, P. K. R., \& Reddy, J. R. (2025, September). GridTrust: A
Secure and Scalable IoT Framework for Vehicle-to-Grid (V2G)
Communication. In~\emph{2025 6th International Conference on Electronics
and Sustainable Communication Systems (ICESC)}~(pp. 1391-1397). IEEE.

{[}35{]} Sarkar, S., Shafaei, S., Jones, T. S., \& Totaro, M. W. (2025).
Secure communication in drone networks: A comprehensive survey of
lightweight encryption and key management
techniques.~\emph{Drones},~\emph{9}(8), 583.

{[}36{]} Lai, C., Ma, J., Wang, X., Zhou, H., \& Zheng, D. (2025). A
novel authentication and key agreement scheme for in-vehicle
networks.~\emph{IEEE Transactions on Vehicular Technology}.

{[}37{]} Venugopal, A., Yogi, K. S., Tamilselvi, M., Ganapathi, R., Rao,
P. V. V., \& Muniyandy, E. (2026). Blockchain-Based Data Integrity and
Provenance Tracking System for Environmental Electrochemical Sensor
Network Analytics.~\emph{Analytical Letters}, 1-22.

{[}38{]} Khan, M. S. M., Biswas, L. K., Kottur, H. R., Noor, R.,
Varshney, N., Hastings, N., \& Asadizanjani, N. (2025). Toward
standardized vulnerability assessment of advanced packaging against
probing attacks.~\emph{IEEE Design \& Test}.

{[}39{]} Zheng, Y., Boyapally, H., Liu, W., Yang, Y., \& Chang, C. H.
(2025). A Lightweight PUF-Based Secure Group Communication Scheme for
Low Altitude Network With Dynamic Group Membership. IEEE Transactions on
Mobile Computing.

{[}40{]} Casado-GalÃ¡n, A., SÃ¡nchez-Solano, S., Tena-SÃ¡nchez, E.,
Rojas-MuÃ±oz, L. F., Potestad-OrdÃ³Ã±ez, F. E., MartÃ­nez-RodrÃ­guez, M. C.,
\& Acosta-JimÃ©nez, A. J. (2025). Analysis of EM Side-Channel Leakage on
an RO-PUF and Proposed Countermeasures.~\emph{IEEE Transactions on
Dependable and Secure Computing}.

{[}41{]} Pawlik, L., Wilk-Jakubowski, J. L., Grabski, P. T., \&
Wilk-Jakubowski, G. (2025). Securing the Electrified Future: A
Systematic Review of Cyber Attacks, Intrusion and Anomaly Detection, and
Authentication in Electric Vehicle Charging
Infrastructure.~Energies,~18(18), 4847.

{[}42{]} Samanta, S., Ray, B., \& Milenkovic, A. (2025, October).
Analysis of Temperature Effect on SRAM PUF for Low Cost Applications.
In~2025 IEEE Physical Assurance and Inspection of Electronics
(PAINE)~(pp. 1-7). IEEE.

{[}43{]} Yadav, A., Kumar, S., \& Singh, J. (2022). A review of physical
unclonable functions (PUFs) and its applications in IoT
environment.~Ambient Communications and Computer Systems: Proceedings of
RACCCS 2021, 1-13.

{[}44{]} Al-Meer, A., \& Al-Kuwari, S. (2023). Physical unclonable
functions (PUF) for IoT devices.~ACM Computing Surveys,~55(14s), 1-31.

{[}45{]} Zhang, Q., Wu, J., Zhong, H., He, D., \& Cui, J. (2022).
Efficient anonymous authentication based on physically unclonable
function in industrial internet of things.~IEEE Transactions on
Information Forensics and Security,~18, 233-247.

{[}46{]} Alhamarneh, R. A., \& Mahinderjit Singh, M. (2024).
Strengthening internet of things security: Surveying physical unclonable
functions for authentication, communication protocols, challenges, and
applications.~Applied Sciences,~14(5), 1700.

{[}47{]} Shan, X., Yu, H., Chen, Y., \& Yang, Z. (2023). Physical
unclonable function-based lightweight and verifiable data stream
transmission for industrial iot.~IEEE Transactions on Industrial
Informatics,~19(12), 11573-11583.

{[}48{]} Wei, Y., Ma, Y., Wang, R., Xiao, Y., Xie, Z., Dou, X., ... \&
Wang, J. (2026). Wearableâ€Compatible Allâ€Optical Physical Unclonable
Functions With Hybrid Deep Learningâ€Based Authentication.~Laser \&
Photonics Reviews, e02874.

{[}49{]} Chang, P., Duan, T., Li, X., Lv, Y., Hao, Z., Guo, Y., ... \&
Wang, A. (2026). On-chip nonlinear optical physical unclonable function
based on a thin-film lithium niobate array.~Optics Express,~34(2),
1408-1423.

{[}50{]} Cao, R., Wang, Y., Zhao, L., Wang, Z., \& Mei, N. (2026).
Artificial Intelligence Attack-Resilient Physical Unclonable Functions
from Colloidal Nanowire Randomness.~ACS Applied Materials \& Interfaces.

{[}51{]} Pan, L., Wei, Y., Wang, J., \& Ma, X. (2026). Architected
Nanomaterials Powering Optical Physical Unclonable Functions.~Laser \&
Photonics Reviews, e01958.

{[}52{]} Zhang, H., Pan, Y., Zuo, J., \& Zhang, T. (2026, January).
Physically unclonable function (PUF) structure by self-assembly. In~11th
International Symposium on Advanced Optical Manufacturing and Testing
Technologies (AOMATT 2025)~(Vol. 13992, pp. 513-517). SPIE.

{[}53{]} Baghban-Bousari, N., Eric, D., Palau, G., Crespo-Yepes, A.,
Porti, M., Ramon, E., ... \& Nafria, M. (2026). Feasibility of Physical
Unclonable Functions from Pre-stressed Organic Thin Film Transistors for
Secure Microelectronics.~Microelectronic Engineering,~302, 112407.

{[}54{]} Cheon, Y., Kim, H., Kim, J., Lee, J., \& Ye, J. (2026). Random
graphene adlayer morphologies grown on microfaceted Cu surfaces for
physical unclonable functions.~Journal of Vacuum Science \& Technology
B,~44(1).

{[}55{]} Singh, H. (2025). Artificial Intelligence and Robotics
Transforming Industries with Intelligent Automation
Solutions.~\emph{Available at SSRN 5267868}.

{[}56{]} Lajber, K., SzÅ‘lÅ‘si, J., Szekeres, B. J., \& AndÃ³, M. (2025).
Sensor-based measurement system for welding torch position.~IEEE Sensors
Journal.

{[}57{]} Nag, A., Hassan, M. M., Das, A., Sinha, A., Chand, N., Kar, A.,
... \& Alkhayyat, A. (2024). Exploring the applications and security
threats of Internet of Thing in the cloud computing paradigm: A
comprehensive study on the cloud of things.~Transactions on Emerging
Telecommunications Technologies,~35(4), e4897.

{[}58{]} Oladosu, S. A., Ige, A. B., Ike, C. C., Adepoju, P. A., Amoo,
O. O., \& Afolabi, A. I. (2022). Reimagining multi-cloud
interoperability: A conceptual framework for seamless integration and
security across cloud platforms.~Open Access Res J Sci Technol,~4(1),
26.

{[}59{]} Parnianifard, A., Jearavongtakul, S., Sasithong, P., Sinpan,
N., Poomrittigul, S., Bajpai, A., ... \& Wuttisittikulkij, L. (2022).
Digital-twins towards cyber-physical systems: a brief
survey.~Engineering Journal,~26(9), 47-61.

{[}60{]} Kaur, M. J., Riaz, S., \& Mushtaq, A. (2019). Cyber-physical
cloud computing systems and internet of everything. In~Principles of
Internet of Things (IoT) Ecosystem: Insight Paradigm~(pp. 201-227).
Cham: Springer International Publishing.

{[}61{]} Xu, H., Yu, W., Griffith, D., \& Golmie, N. (2018). A survey on
industrial Internet of Things: A cyber-physical systems
perspective.~Ieee access,~6, 78238-78259.

{[}62{]} Mazhar, T., Shahzad, T., Rehman, A. U., \& Hamam, H. (2025).
Integration of smart grid with industry 5.0: applications, challenges
and solutions.~Measurement: Energy,~5, 100031.

{[}63{]} Kim, S., Park, K. J., \& Lu, C. (2022). A survey on network
security for cyber--physical systems: From threats to resilient
design.~IEEE Communications Surveys \& Tutorials,~24(3), 1534-1573.

{[}64{]} Ribas Monteiro, L. F., Rodrigues, Y. R., \& Zambroni de Souza,
A. C. (2023). Cybersecurity in cyber--physical power systems. Energies,
16(12), 4556.

{[}65{]} Antonioli, D., \& Tippenhauer, N. O. (2015, October). MiniCPS:
A toolkit for security research on CPS networks. In~Proceedings of the
First ACM workshop on cyber-physical systems-security and/or
privacy~(pp. 91-100).

{[}66{]} Adam, M., Hammoudeh, M., Alrawashdeh, R., \& Alsulaimy, B.
(2024). A survey on security, privacy, trust, and architectural
challenges in IoT systems.~IEEE Access,~12, 57128-57149.

{[}67{]} Jaime, F. J., MuÃ±oz, A., RodrÃ­guez-GÃ³mez, F., \& Jerez-Calero,
A. (2023). Strengthening privacy and data security in biomedical
microelectromechanical systems by IoT communication security and
protection in smart healthcare.~Sensors,~23(21), 8944.

{[}68{]} Knapp, E. D. (2024).~Industrial Network Security: Securing
critical infrastructure networks for smart grid, SCADA, and other
Industrial Control Systems. Elsevier.

{[}69{]} Wang, Z., Xie, W., Wang, B., Tao, J., \& Wang, E. (2021). A
survey on recent advanced research of CPS security.~Applied
Sciences,~11(9), 3751.

{[}70{]} Dhavlle, A., Hassan, R., Mittapalli, M., \& Dinakarrao, S. M.
P. (2021, May). Design of hardware trojans and its impact on cps
systems: A comprehensive survey. In~2021 IEEE International Symposium on
Circuits and Systems (ISCAS)~(pp. 1-5). IEEE.

{[}71{]} Alsabbagh, W., \& LangendÃ¶rfer, P. (2023). Security of
programmable logic controllers and related systems: Today and
Tomorrow.~IEEE Open Journal of the Industrial Electronics Society,~4,
659-693.

{[}72{]} Kitchin, R., \& Dodge, M. (2020). The (in) security of smart
cities: Vulnerabilities, risks, mitigation, and prevention. In~Smart
cities and innovative Urban technologies~(pp. 47-65). Routledge.

{[}73{]} Alsuwaidi, N., Alharmoodi, N., \& Al Hamadi, H. (2024,
October). Securing Smart Grid Infrastructures: Challenges, Defense
Mechanisms, and Future Directions. In~2024 IEEE Future Networks World
Forum (FNWF)~(pp. 933-940). IEEE.

{[}74{]} Amin, M., El-Sousy, F. F., Aziz, G. A. A., Gaber, K., \&
Mohammed, O. A. (2021). CPS attacks mitigation approaches on power
electronic systems with security challenges for smart grid applications:
A review.~Ieee Access,~9, 38571-38601.

{[}75{]} Kure, H. I., Islam, S., \& Razzaque, M. A. (2018). An
integrated cyber security risk management approach for a cyber-physical
system.~Applied Sciences,~8(6), 898.

{[}76{]} Coburn, J., Ravi, S., Raghunathan, A., \& Chakradhar, S. (2005,
September). Seca: security-enhanced communication architecture. In
Proceedings of the 2005 international conference on Compilers,
architectures and synthesis for embedded systems (pp. 78-89).

{[}77{]} Fournaris, A. P., \& Sklavos, N. (2014). Secure embedded system
hardware design--A flexible security and trust enhanced
approach.~Computers \& Electrical Engineering,~40(1), 121-133.

{[}78{]} Maene, P., GÃ¶tzfried, J., De Clercq, R., MÃ¼ller, T., Freiling,
F., \& Verbauwhede, I. (2017). Hardware-based trusted computing
architectures for isolation and attestation.~IEEE Transactions on
Computers,~67(3), 361-374.

{[}79{]} Rahman, M. H. (2024). A Comprehensive Survey on
Hardware-Software co-Protection against Invasive, Non-Invasive and
Interactive Security Threats.~Cryptology ePrint Archive.

{[}80{]} Khan, M. S. M. (2025).~Physical Attack Resilience and
Authentication Strategies for Multi-Chiplet Integrated Circuits (IC)
With Advanced Packaging~(Doctoral dissertation, University of Florida).

{[}81{]} Yang, K., Blaauw, D., \& Sylvester, D. (2017). Hardware designs
for security in ultra-low-power IoT systems: An overview and survey.
IEEE Micro, 37(6), 72-89.

{[}82{]} Hassija, V., Chamola, V., Saxena, V., Jain, D., Goyal, P., \&
Sikdar, B. (2019). A survey on IoT security: application areas, security
threats, and solution architectures. IEEE access, 7, 82721-82743.

{[}83{]} Fazeldehkordi, E., \& GrÃ¸nli, T. M. (2022). A survey of
security architectures for edge computing-based IoT. IoT, 3(3), 332-365.

{[}84{]} Sellami, Y. (2024). Secure data management in an IoT-Fog/Edge
computing architecture (Doctoral dissertation, UniversitÃ© Polytechnique
Hauts-de-France).

{[}85{]} Kumar, S., Kumar, D., Dangi, R., Choudhary, G., Dragoni, N., \&
You, I. (2024). A review of lightweight security and privacy for
resource-constrained IoT devices. Computers, Materials and Continua,
78(1), 31-63.

{[}86{]} Tehranipoor, M., Pundir, N., Vashistha, N., \& Farahmandi, F.
(2023).~Hardware security primitives. Switzerland: Springer.

{[}87{]} Sharma, G. (2025). A survey on lightweight hardware security
using physically unclonable functions for IoT devices.~Peer-to-Peer
Networking and Applications,~18(6), 315.

{[}88{]} Kumar, V., \& Paul, K. (2023). Device fingerprinting for
cyber-physical systems: A survey. ACM Computing Surveys, 55(14s), 1-41.

{[}89{]} Paral, Z., \& Devadas, S. (2011, June). Reliable and efficient
PUF-based key generation using pattern matching. In~2011 IEEE
international symposium on hardware-oriented security and trust~(pp.
128-133). IEEE.

{[}90{]} Shin, C. (2016). Variation-aware advanced CMOS devices and SRAM
(Vol. 56). Dordrecht: Springer Netherlands.

{[}91{]} Johnson, D. S., Li, W., Gordon, D. B., Bhattacharjee, A.,
Curry, B., Ghosh, J., ... \& Liu, X. S. (2008). Systematic evaluation of
variability in ChIP-chip experiments using predefined DNA
targets.~Genome research,~18(3), 393-403.

{[}92{]} Mishra, A. (2025).~ENHANCING THE SECURITY SCALABILITY OF
ARBITER PUFS USING MEMORY-BASED WEAK PUFS~(Doctoral dissertation, Purdue
University Graduate School).

{[}93{]} Herkle, A. (2023). Techniques to enhance the reliability of
delay-based physical unclonable functions.

{[}94{]} Yao, Y., Kim, M., Li, J., Markov, I. L., \& Koushanfar, F.
(2013, March). ClockPUF: Physical Unclonable Functions based on clock
networks. In~2013 Design, Automation \& Test in Europe Conference \&
Exhibition (DATE)~(pp. 422-427). IEEE.

{[}95{]} Syed, Y. (2026).~A Novel Configurable Ring Oscillator Physical
Unclonable Function Design for Enhanced IoT
Security~(Master\textquotesingle s thesis, The Catholic University of
America).

{[}96{]} Maes, R. (2013). Physically Unclonable Functions:
Constructions, Properties and Applications. Springer.

{[}97{]} Avvaru, S. S., Zhou, C., Satapathy, S., Lao, Y., Kim, C. H., \&
Parhi, K. K. (2016, March). Estimating delay differences of arbiter PUFs
using silicon data. In~2016 Design, Automation \& Test in Europe
Conference \& Exhibition (DATE)~(pp. 543-546). IEEE.

{[}98{]} Ferens, M., Dushku, E., \& Kosta, S. (2024). When Random is
Bad: Selective CRPs for Protecting PUFs against Modeling Attacks.~IEEE
Transactions on Computer-Aided Design of Integrated Circuits and
Systems.

{[}99{]} Anandakumar, N. N., Hashmi, M. S., \& Sanadhya, S. K. (2020).
Efficient and lightweight FPGA-based hybrid PUFs with improved
performance. Microprocessors and Microsystems, 77, 103180.

{[}100{]} Singh, H. (2024). AI-Enabled Hardware Security Approach for
Aging Classification and Manufacturer Identification of SRAM PUFs.

{[}101{]} Merli, D., Stumpf, F., \& Eckert, C. (2010, October).
Improving the quality of Ring-Oscillator PUFs on FPGAs. In~Proceedings
of the 5th workshop on embedded systems security~(pp. 1-9).

{[}102{]} Hemavathy, S., \& Bhaaskaran, V. K. (2023). Arbiter PUF---A
review of design, composition, and security aspects.~IEEE Access,~11,
33979-34004.

{[}103{]} Anandakumar, N. N., Hashmi, M. S., \& Sanadhya, S. K. (2022).
Design and analysis of FPGA-based PUFs with enhanced performance for
hardware-oriented security.~ACM Journal on Emerging Technologies in
Computing Systems (JETC),~18(4), 1-26.

{[}104{]} Ahsan, S. M., Hossain, T., Hasan, M. S., \& Hoque, T. (2023,
April). Resistive ram-based puf: Challenges and opportunities. In~2023
IEEE 16th Dallas Circuits and Systems Conference (DCAS)~(pp. 1-6). IEEE.

{[}105{]} Whig, P., Batra, I., Yathiraju, N., \& Jain, S. N. (2025).
Blockchain for Hardware Security and Trust. In~Hardware Security:
Challenges and Solutions~(pp. 27-49). Cham: Springer Nature Switzerland.

{[}106{]} Chen, H., \& Babar, M. A. (2024). Security for machine
learning-based software systems: A survey of threats, practices, and
challenges.~ACM Computing Surveys,~56(6), 1-38.

{[}107{]} Kornaros, G. (2022). Hardware-assisted machine learning in
resource-constrained IoT environments for security: review and future
prospective. IEEE Access, 10, 58603-58622.

{[}108{]} Wang, J., Wang, J., Fan, C., Yan, F., Cheng, Y., Zhang, Y.,
... \& Hu, H. (2023). SvTPM: SGX-based virtual trusted platform modules
for cloud computing.~IEEE Transactions on Cloud Computing,~11(3),
2936-2953.

{[}109{]} Asif, R., Ghanem, K., \& Irvine, J. (2020). Proof-of-puf
enabled blockchain: Concurrent data and device security for
internet-of-energy.~Sensors,~21(1), 28.

{[}110{]} Pandey, S., \& Bhushan, B. (2024). Recent Lightweight
cryptography (LWC) based security advances for resource-constrained IoT
networks.~Wireless Networks,~30(4), 2987-3026.

{[}111{]} Pundir, N. K. (2017). Design of a hardware security puf immune
to machine learning attacks (Master\textquotesingle s thesis, University
of Toledo).

{[}112{]} Prakash, K. (2023).~Building modeling resistant Physically
Unclonable Functions (PUFs) using Adversarial Machine Learning. McGill
University (Canada).

{[}113{]} Zhang, Y., Li, B., Liu, B., \& Chang, J. (2024). Building PUF
as a service: Distributed authentication and recoverable data sharing
with multidimensional CRPs security protection.~IEEE Internet of Things
Journal,~11(10), 17301-17316.

{[}114{]} Ebrahimabadi, M., Younis, M., \& Karimi, N. (2021). A
PUF-based modeling-attack resilient authentication protocol for IoT
devices.~IEEE Internet of Things Journal,~9(5), 3684-3703.

{[}115{]} Oun, A. (2022).~Hardware Security Design, and Vulnerability
Analysis of FPGA based PUFs to Machine Learning and Swarm Intelligence
based ANN Algorithm Attacks~(Doctoral dissertation, University of
Toledo).

{[}116{]} Koeberl, P., KocabaÅŸ, Ãœ., \& Sadeghi, A. R. (2013, March).
Memristor PUFs: a new generation of memory-based physically unclonable
functions. In~2013 Design, Automation \& Test in Europe Conference \&
Exhibition (DATE)~(pp. 428-431). IEEE.

{[}117{]} Cao, H., \& Wang, F. (2023). An overview of complex
instability behaviors induced by nonlinearity of power electronic
systems with memristive load.~Energies,~16(6), 2528.

{[}118{]} Oduro-Antwi, M., Nguyen, D., \& Sood, K. (2026). Physically
unclonable functions (PUF)-based IoT security: challenges and
opportunities.~Internet of Things Security, 201-217.

{[}119{]} Alotaibi, A., \& Rassam, M. A. (2023). Adversarial machine
learning attacks against intrusion detection systems: A survey on
strategies and defense.~Future Internet,~15(2), 62.

{[}120{]} Lyu, Y., \& Mishra, P. (2018). A survey of side-channel
attacks on caches and countermeasures.~Journal of Hardware and Systems
Security,~2(1), 33-50.

{[}121{]} Shrivastwa, R. R. (2023).~Enhancements in Embedded Systems
Security using Machine Learning~(Doctoral dissertation, Institut
Polytechnique de Paris).

{[}122{]} Taghizadeh, S., Taghizadeh, M., Taghizadeh, P., Kamaly, A., \&
Emamghorashi, S. A. (2016). Design of a New LC VCO using Active
Inductor. International Journal of Computer Sciences and Engineering,
4(12), 27-30.

{[}123{]} Taghizadeh, M., Taghizadeh, P., Taghizadeh, S., Kamaly, A., \&
Emamghorashi, S. A. (2016). Design of a Novel Ring VCO with low Phase
Noise and High frequency range. International Journal of Computer
Sciences and Engineering, 4(12), 8-12.

{[}124{]} Pishdast, H., Kalva, H., \& Tye, D. (2025). AI-Enabled Smart
Camera Traps for Wildlife Monitoring in African Ecosystems. In
Proceedings of the 2025 International Conference on Information
Technology for Social Good (GoodIT \textquotesingle25), September 3-5,
2025, Antwerp, Belgium. ACM. https://doi.org/10.1145/3748699.3749780

\end{document}